# A Fast Heuristic for Gateway Location in Wireless Backhaul of 5G Ultra-Dense Networks

Mital Raithatha[1], Aizaz U. Chaudhry[1], Senior Member, IEEE, Roshdy H.M. Hafez[1], Life Senior Member, IEEE, and John W. Chinneck[1]

[1]Department of Systems and Computer Engineering, Carleton University, Ottawa, ON K1S 5B6, Canada

Corresponding author: Mital Raithatha (e-mail: mitalraithatha@cmail.carleton.ca)

**ABSTRACT** In 5G Ultra-Dense Networks, a distributed wireless backhaul is an attractive solution for forwarding traffic to the core. The macro-cell coverage area is divided into many small cells. A few of these cells are designated as gateways and are linked to the core by high-capacity fiber optic links. Each small cell is associated with one gateway and all small cells forward their traffic to their respective gateway through multi-hop mesh networks. We investigate the *gateway location problem* and show that finding near-optimal gateway locations improves the backhaul network capacity. An exact *p*-median integer linear program is formulated for comparison with our novel K-GA heuristic that combines a Genetic Algorithm (GA) with *K*-means clustering to find near-optimal gateway locations. We compare the performance of K-GA with six other approaches in terms of average number of hops and backhaul network capacity at different node densities through extensive Monte Carlo simulations. All approaches are tested in various user distribution scenarios, including uniform distribution, bivariate Gaussian distribution, and cluster distribution. In all cases K-GA provides near-optimal results, achieving average number of hops and backhaul network capacity within *2%* of optimal while saving an average of *95%* of the execution time.

## I. INTRODUCTION

The recent introduction of 5G networks around the world induced the development and deployment of many wireless services, such as, ultra-high-definition video streaming, augmented reality, sophisticated on-line video gaming, security applications, intelligent farming, and connected vehicles. 5G has three stated objectives: (1) support for Enhanced-Mobile Broadband services (eMBB), (2) support for ultra-Reliable and Low Latency services (uRLL), and (3) support for massive Machine Type Communications (mMTC). This paper addresses the first stated objective where 5G networks aim to increase the data rate by as much as two orders of magnitude [1].

Expanding a wireless network's capacity by two orders of magnitude or more is an ambitious target, but extensive research and development efforts have put this target within reach. In the quest for much faster data transmission, 5G networks are deploying several technologies to improve network capacity and spectrum efficiency. One of these technologies is *massive MIMO* (multiple-input multiple-output), which helps improve the channel capacity and signal strength by employing multiple antennas for transmission and reception [2]. Another important technology is the *millimeter wave (mm-wave) band*. 5G is set to exploit the massive spectrum space available at higher frequencies. The mm-wave frequencies are expected to provide hundreds of megahertz of bandwidth to meet the requirements of higher data rates [3]. In addition to massive MIMO and mm-wave, *network densification* is the third and most promising approach to handle higher spectrum demands in crowded venues. Densification of crowded cells increases network capacity in terms of bits/sec/Hz/unit area. The basic approach is to make the network as dense as possible by deploying a large number of access nodes within a coverage area. These access nodes are referred to as "small cells". Cell densification improves link quality and significantly increases network capacity [4]. The combination of wider RF bandwidth, massive MIMO, and deployment of many small cells gives rise to what is now known as an *Ultra-Dense Network* (UDN) [5], in which cells have a very high data rate per unit area.

While cell densification using a UDN increases the capacity in terms of bits/sec/Hz/unit area, it complicates the backhauling problem. "Backhaul" refers to the links between access points and the core network. An obvious solution to the backhaul problem is to directly connect each small cell to the core using fiber optic or any broadband cabling, but this is costly and cumbersome. On the other hand, a wireless backhaul solution is flexible and cost-effective. In particular, the use of mm-wave bands provides the spectrum resources needed to connect small cells to gateways [6]. The idea of using wireless links to facilitate backhauling received special attention in 3GPP Release 16 [7]. One of the key novelties in Release 16 is to integrate access and backhaul using small cells called *Integrated Access and Backhaul* (IAB) nodes. An IAB node is a miniature base station that communicates with User Equipment (UEs) on a given frequency and delivers UE traffic to the core network on a different frequency as part of the backhaul network. This technology has gained much attention in the industry because IAB is seen as a cost efficient and convenient solution [8].

This paper explores the use of IAB nodes (or small cells) in multi-hop wireless networks that carry the access traffic to the core. We use the term "distributed wireless backhaul" to imply that $N$ small cells are clustered into groups, each group is associated with a *gateway* (GW) by means of a multi-hop mesh network, and all connections among small cells (including gateways) use mm-wave links. Our network model is based on the IAB architecture in Release 16, where an IAB node is referred to as a small cell and an IAB donor is called a gateway. IAB nodes (small cells) and donors (gateways) collectively form a multi-hop backhaul architecture. Proper backhaul network design is crucial for maximizing the *backhaul network capacity* (BNC).

A well-designed wireless-based backhaul network should deliver all access traffic intercepted at all small cells to the core using minimum spectral resources. There are two important considerations: (1) *Bottleneck avoidance*: we must ensure that small cell to small cell link capacities are large enough to accommodate the accumulated traffic volume, especially near the gateways; and (2) *Redundancy minimization*: information bits intercepted at small cells will be transmitted several times during their journey to the gateways. Minimizing the *Average Number of Hops* (ANH) throughout the network is crucial in addressing these two considerations and in improving the backhaul network spectral efficiency (bits/s/Hz/unit area) [9].

The *gateway location problem* (GLP) studied in this paper involves finding gateway locations ($M$ out of $N$ small cells) such that BNC is maximized. This means finding the smallest ANH over every possible combination of $M$ out of $N$ selections, which is combinatorially explosive. For example, if $N = 400$ and $M = 4$, then there are more than a billion possible solutions. The GLP is similar to the *p*-median problem, which is known to be NP-Hard [10]. We have formulated the *p*-median problem as an *Integer Linear Program* (ILP) to find provably optimal gateway locations, but as the size of the problem instance increases, the ILP becomes too large to solve in a reasonable amount of time. For this reason, we explore heuristic methods based on *Artificial Intelligence* (AI) and *Machine Learning* (ML) to find GW locations and to associate small cells to them.

Our new *K-GA* algorithm combines (i) ML's *K*-means clustering algorithm with (ii) AI's genetic algorithm and (iii) Dijkstra's shortest path algorithm from operations research. *Genetic Algorithms* (GAs) are stochastic optimization heuristics inspired by biological evolution [11]. The well-known *K*-means and *K*-medoids clustering are unsupervised machine learning algorithms, which are simple partition-based algorithms where $k$ clusters create $k$ centroids or $k$ medoids [12][13]. Dijkstra's shortest path algorithm [14] is used to associate small cells to GWs. Preliminary work on the algorithm appeared in [15]. The main contribution of the work reported here is as follows:

- The development of a new heuristic algorithm (K-GA), which provides locations for GWs within *2%* of the optimal locations in all small cell distribution scenarios. The proposed heuristic is much faster than the exact integer programming method.
- We investigate various network topologies under different small cell distribution scenarios (uniform, bivariate Gaussian, and cluster distributions) to assess the performance of the proposed K-GA heuristic.
- Extensive Monte Carlo based simulations compare the proposed K-GA with *K*-means, GA, *K*-medoids, a baseline approach, and a hybrid of *K*-medoids and GA, in terms of ANH and BNC.

The rest of this paper is organized as follows. After providing the background information on the *p*-median problem, Genetic Algorithms, *K*-means clustering algorithm, *K*-medoids clustering algorithm, and Dijkstra's shortest path algorithm, Section II reviews the relevant literature. Section III describes the network model of the distributed multi-hop wireless backhaul. Section IV gives the problem formulation in terms of BNC and Section V formulates the mathematical model of the GLP. The K-GA heuristic is detailed in Section VI. Section VII describes the experimental methodologies, and the results are presented in Section VIII. Section IX concludes the paper with an outlook on future work.

## II. BACKGROUND AND RELATED WORK

### A. BACKGROUND
#### 1) *P*-MEDIAN PROBLEM
The *p*-median problem is in the larger class of *minisum location–allocation problems* [16]. The goal is to locate *p* facilities (medians) to minimize the total weighted distance between the median points and the demand points. Hakimi [17] introduced the median location problem on graphs in 1964. Several methods have been developed for general networks. If the basic graph of the network is a tree, then the *p*-median problem can be solved with known algorithms in polynomial time.

*a: COMPLEXITY OF P-MEDIAN PROBLEM*
Matula and Kolde [18] provided an algorithm in 1976 with complexity $O(N^3p^2)$ for locating the *p*-medians of a tree where *p* is the number of facilities to select out of *N* facilities and $p > 1$. In 1979, Kariv and Hakimi [19] proved that the *p*-median location problem on a general network is NP-hard. In addition, they also investigated the *p*-median problem on tree graph networks and designed an algorithm with complexity $O(N^2p^2)$. A new algorithm was designed in 1982 for trees with complexity $O(N^3p)$ by Hsu [20]. Around 15 years later, Tamir [21] improved the time complexity on tree networks to $O(N^2p)$. In 2005, Benkoczi and Bhattacharya [22] designed an algorithm for trees with $O(N^2 log^{p+2})$ runtime.

2) GENETIC ALGORITHMS
A GA is parallel in nature, which improves speed when applied to the *p*-median problem [23]. GAs are stochastic, efficient, and easily manageable for complex problems and have been widely used in analyzing data, integrating information, and using the resulting insights to improve decision making [24][25]. In the fields of neural networks, computer science, machine learning, artificial life and others, GAs are used as a stochastic search and optimization heuristic [26][27].

GAs search for a suitable solution through evolution and randomness. A solution is represented as a "chromosome" string. A population of chromosomes is initially generated randomly. These have an associated fitness score, which affects their probability of being selected for the mating pool. Pairs of chromosomes are selected from the mating pool for the crossover operation, which creates new chromosomes by swapping their ends at a random crossover point. These chromosomes are then subjected to the mutation operation, which changes the values of some elements randomly. These operations produce a new population of chromosomes (generation), and over a series of generations, better solutions are evolved [28][29].

*a: COMPLEXITY OF GENETIC ALGORITHMS*
The computational complexity of a genetic algorithm is well studied [30][31][32][33]. It depends mainly on the problem size, the fitness function, and parameters such as the selection probability, mutation probability, number of chromosomes, etc. In our problem, the GA's complexity is $O(NGn_{pop})$, where *N* is the number of nodes or data points, *G* is the number of generations, and $n_{pop}$ is the number of chromosomes [34]. Assume that the mutation probability ($P_m$) and crossover probability ($P_c$) for single point crossover are less than 1. The selection process completes its operation in $n_{pop}$ operations at each iteration. Single point crossover exchanges the values in $O(NP_cn_{pop}/2)$ time and the mutation process changes elements in $O(NP_mn_{pop})$ time, where $P_c<<1$ and $P_m<<1$, and this reduces the complexity to $O(Nn_{pop})$ in a single generation. So, the overall complexity of a genetic algorithm for *G* generations is $O(NGn_{pop})$.

3) K-MEANS CLUSTERING ALGORITHM
*K*-means clustering is a popular and widely used machine learning algorithm for data mining across different disciplines. It is used to process large amounts of unstructured data [35][36]. The main goal is to divide the *N* data points into *M* clusters so that the within-cluster total squared Euclidean distance is minimized [37].

*a: COMPLEXITY OF K-MEANS CLUSTERING ALGORITHM*
The computational complexity of *K*-means clustering is $O(NMI)$ [38], where *N* is the number of data points (number of small cells in our case), *M* is the number of clusters, and *I* is the number of iterations. Generally, $M << N$, and the computational complexity of *K*-means reduces to $O(NI)$.

4) K-MEDOIDS CLUSTERING ALGORITHM
The *K*-medoids clustering algorithm is similar to *K*-means clustering, but the cluster center, called a medoid, must be a member of that cluster. Unlike *K*-means, this algorithm returns medoids that are actual nodes. It uses *partitioning around medoids* (PAM) [39] and proceeds in two steps:
- *Build*: For cluster center initialization, *M* nodes out of *N* are selected randomly as medoids. Then, *M* clusters are constructed by assigning each node to the nearest medoid based on squared Euclidean distance.
- *Swap*: Within each cluster, each node (small cell in our case) is tested as a potential medoid by checking whether the sum of within-cluster distances gets smaller using that node as the medoid. If so, the node is defined as a new medoid.

*K*-medoids clustering iterates through the build and swap steps until the medoids do not change.

*a: COMPLEXITY OF K-MEDOIDS CLUSTERING ALGORITHM*
The computational complexity of *K*-medoids is $O(M(N-M)^2I)$ [40], where *M* is the number of clusters or gateways, *N* is the number of nodes (number of small cells in our case), and *I* is the number of iterations. Generally, $M << N$, and the computational complexity of *K*-medoids reduces to $O(N^2I)$. *K*-means is more efficient compared to *K*-medoids in cases where the number of data points (small cells) is large.

5) DIJKSTRA'S SHORTEST PATH ALGORITHM
Dijkstra's algorithm is a well-known way to find the shortest path between two points in a given network [14]. A variant finds the shortest paths from a source node to all other nodes for a given graph $G\,(V,\,E)$ where $V$ is the set of nodes and $E$ is the set of edges. Each of the edges in $E$ has a weight, which represents the length of the edge in terms of hops or distance. See e.g. [41][42][43] for details and pseudocode.

*a: COMPLEXITY OF DIJKSTRA'S SHORTEST PATH ALGORITHM*

The computational complexity of Dijkstra's algorithm for a single source shortest path is $O(N^2)$ [44][45], where $N$ is the number of nodes. So, for $N$ sources of shortest path trees, the computational complexity becomes $O(N^3)$.

*B. RELATED WORK*

Two popular backhaul architectures have been proposed in the literature [9][46]: "centralized" and "distributed". In the centralized architecture, the small cells are connected to a single point (usually the site of a macro-cell) through which all traffic is backhauled to the core by fiber optic cables. In the distributed architecture, the small cells are clustered into groups. Each small cell connects to a gateway and gateways connect to the core via fiber. Most reported work compares the centralized and distributed schemes. The distributed architecture is shown to have better results in terms of capacity and energy efficiency [9], and the distributed architecture achieves higher throughput [46].

In the recent 3GPP Release 16, mm-wave is an acceptable backhaul solution for small cell networks [7], and numerous studies [47][48][49] on wireless backhaul technologies highlight mm-wave wireless backhaul as the most acceptable solution for 5G communications. [47] investigated the advantages and disadvantages of mm-wave and free space optics (FSO) for fronthaul/backhaul links and showed mm-wave to be better than FSO in terms of energy efficiency and availability. A mm-wave backhaul-based massive MIMO scheme is proposed in [48] for 5G ultra dense networks. It is shown that mm-wave can be easily merged with Massive MIMO by deploying a large number of antennas in the wireless backhaul network. [49] evaluated the advantages such as low latency and high quality of service in a mm-wave wireless backhaul.

There have been several studies on different aspects of IAB networks in 5G ultra-dense scenarios. For instance, the IAB node (or small cell) placement problem [50] [51] [52] [53], interference management [54] [55] [56] [57] [58] [59], resource allocation [60] [61], the IAB donor (or gateway) placement problem [62] [63] [64] [65] [66], and mobility management [67] [68] [69] have been studied in the context of 5G ultra-dense networks. The focus in this paper is on the gateway location problem in 5G ultra-dense networks.

Several works have been proposed for the GLP in scenarios such as Wireless Mesh Networks, Wireless Sensor Networks, and Satellite Networks. The work in [70] focused on gateway placement in 5G satellite hybrid networks for improving network reliability. An optimal enumeration algorithm and a cluster-based approximate placement algorithm are applied for the GLP. [71] proposed the method of multiple surface gateways positioning in underwater sensor networks. [72][73][74] focused on the GLP in wireless mesh networks, showing that a GA works better than other algorithms for optimizing the locations of gateways. [72] used a genetic algorithm for minimizing the variance in hops count between each internet gateway and its associated mesh router in the network. [73] implemented a genetic algorithm and a simulated annealing algorithm for optimization in WMN for improving performance in terms of cost and quality of service. In [74], the comparative analysis shows that a genetic algorithm performs best among all combinatorial algorithms.

There is little work on the GLP in UDNs. In [62], a fiber backhaul is compared with a wireless backhaul solution in the IAB architecture. The performance of the IAB network is evaluated for applications such as the 3GPP HTTP model. A subset of the IAB nodes are assumed to be IAB donors (or gateways) and no gateway placement strategy is involved, which may result in overloading some IAB-donors or increasing the number of hops. In [63] the authors design a wireless backhaul network planner to reduce the cost of the deployment. The main objective of the algorithm was to maximize the overall coverage while minimize the number of gateways. [64] focused on the joint selection of cluster heads and number of base station antennas to maximize the overall system throughput. [65] concentrates on maximizing the wireless BNC, but optimally selecting gateways was not investigated. [66] proposed a solution for the gateway placement to increase energy efficiency. A comparative summary of our proposed scheme with the existing work on GLP in 5G ultra-dense networks is presented in Table 1.

Unlike [62], [63], [64], [65], [66], which address GLP in a different way, our work aims to fill the research gap found in previous studies by proposing the K-GA heuristic for locating a given number of gateways such that the average number of hops from small cells to gateways is minimized and backhaul network capacity is maximized in an efficient way.

This paper is a significant extension of work originally reported in [15]. We develop an ILP formulation for the *p-median* problem to find optimal gateway locations and compare the optimal solutions with the proposed K-GA heuristic. We have also added the *K-medoid algorithm* for comparison with K-GA to validate its performance. Our earlier work tested a uniform distribution scenario, which is the standard and simplest assumption about user distribution within the coverage area. When users are uniformly distributed, we also assume that the small cells will be uniformly distributed. This is a mathematically simple model, but rarely happens in the real world. In this paper we also simulate more realistic distributions of users and small cells such as the bivariate Gaussian distribution and a cluster distribution [75]. We have also analyzed the computational complexity of all approaches and assessed execution time of K-GA in comparison with the optimal ILP approach.

TABLE 1
SUMMARY OF EXISTING SOLUTIONS ON GLP IN 5G UDNs

| Paper | Objective | Strength | Weakness | Performance Metrics | | |
|---|---|---|---|---|---|---|
| | | | | Mathematical Solution | Computational Complexity | Comparison Methods |
| [62] | IAB networks are investigated to examine the deployment cost of 5G mm-wave networks. | Compared the performance of different applications using three different backhaul scenarios. | • No gateway selection strategy is involved.<br>• Benefits decrease for more congested networks. | Not available | Not available | Not available |
| [63] | Wireless backhaul network planner is designed to maximize the overall coverage. | Minimized the number of installed fiber connections in a wireless backhaul network. | • Priority is given to gateway density by avoiding the criteria for number of hops.<br>• Relaxed some of the constraints to find optimal solution. | Used a greedy approach to approximate the optimal solution and provided upper bound by removing integer constraints. | Not available | Not available |
| [64] | Proposed dynamic selection of cluster heads to maximize the overall system throughput. | Joint selection of cluster heads and antenna partitioning to achieve higher throughput. | • Total system throughput of proposed solution decreases with increase of small cells.<br>• Relaxed the problem to obtain optimal solution. | Proposed polynomial time algorithm to obtain the optimal solution. | $O(\|N\| \log(\|N\|)) + O(\|N\|^2 \log \|N\|)$ | Compared solution with centralized and hybrid architectures but not with other heuristics or methods. |
| [65] | Wireless backhaul network capacity and energy efficiency are analyzed based on multi-hop wireless networks. | Uniform and Poisson distribution scenarios were explored. | Proposed scheme was not compared with any other approaches or with the optimal solution. | Not available | Not available | Not available |
| [66] | Cost efficiency of 5G wireless backhaul networks is analyzed by optimizing gateway deployment and wireless backhaul routing schemes. | Addressed the optimal deployment of GWs in 5G wireless backhaul networks. | Proposed solution was divided in long time scale and short time scale which make joint optimization algorithm very complex. | Not available | $O(N^4)$ | Compared proposed solution with Bellman Ford and shortest path algorithms. |
| Our Paper | Gateway location problem is studied which involves finding gateway locations such that ANH is minimized, and BNC is maximized. | Addressed the combinatorically explosive GLP, obtained the optimal solution, developed a new fast heuristic and compared with other heuristic approaches under various distribution scenarios. | Does not consider access part of the network. | Used CPLEX solver to find the optimal solution. | $O(NI + N^2 G n_{pop})$ | Compared K-GA with Genetic Algorithms, *K*-means, *K*-medoids, KM-GA and baseline approach. |

Our approach directly tackles the GLP in a 5G multi-hop ultra-dense scenario to increase the wireless backhaul network capacity. K-GA provides promising near-optimal results in all distribution scenarios and achieves better results than other heuristics. It is significantly faster than the optimum ILP.

### III. NETWORK MODEL
We consider a circular UDN where the coverage area is divided into $N$ small cells. $M$ out of $N$ cells must be designated as gateways. The gateways are connected to the core network by fiber optic links with very high capacity. The remaining ($N$-$M$) cells are grouped into $M$ clusters with each cluster served by one gateway. Each small cell connects to its serving gateway either directly or through multiple hops within the cluster. We further assume that small cells use mm-wave wireless links to connect with each other or with the gateway [65]. Our work does not consider the access part of the network, and only focuses on the wireless backhaul for the 5G ultra dense network. The overall model is shown in Fig.1.

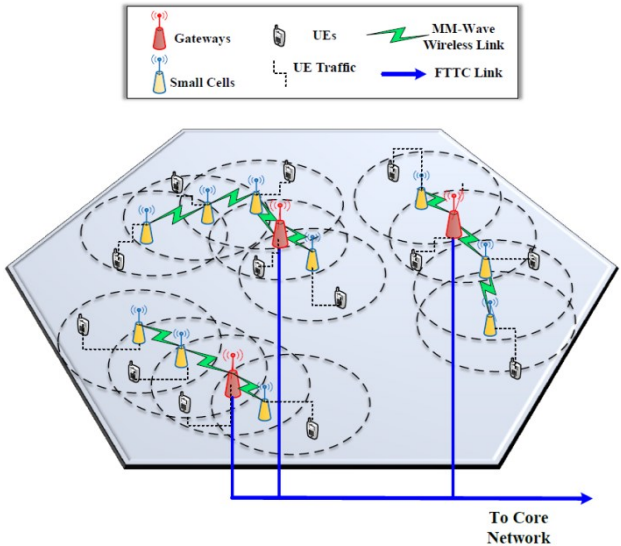

**FIGURE 1.** Network Model

The main assumptions for this model are listed below:
- Each small cell has a fixed circular backhaul coverage area.
- The backhaul coverage areas of small cells are equal.
- Two small cells must be in the transmission range of each other to communicate.
- Each small cell has the same capacity $W_S$, which is larger than the incoming access traffic.
- The capacity of a gateway is $W_G$, which is the capacity of the fiber link between the GW and the core.
- Any small cell can be selected as a gateway for forwarding backhaul traffic to the core network.

### IV. PROBLEM FORMULATION
Our main goal is to choose $M$ gateways out of $N$ small cells such that the average number of hops in the entire coverage area is minimized, and the spectral utilization efficiency and thereby the backhaul network capacity of the wireless backhaul network is maximized, where $M$ is given. Each small cell has wireless links to its neighbor with capacity $W_S$. The capacity of a gateway is denoted by $W_G$. A formula for the backhaul network capacity of the UDN is given below [65]:

$$C(M, N) = \frac{\min(N \cdot W_S, \ M \cdot (W_G - W_S))}{\min(\overline{Y(M,N)})} + M \cdot W_S \qquad (1)$$

where $\min(\overline{Y(M,N)})$ represents the "minimum average number of hops from a small cell to its associated gateway". The "average" is taken over the entire backhaul network.

$$\overline{Y(M,N)} = \frac{1}{N}\sum_{i=1}^{N} Y_i(M,N), \qquad (2)$$

where $Y_i(M,N)$ is the number of hops between $i^{th}$ small cell and its associated gateway, $i = 1,\ldots,N$, and $M \geq 1$.

In developing this formula, the authors maximized the total number of bits generated at all small cells and successfully delivered to all gateways within a period $T$, and took the limit when $T$ tended to infinity. Each information bit was counted once. For a given spectral bandwidth, the information throughput is achieved by maximizing the number of simultaneous transmissions (represented by the numerator of the first term) divided by the average number of hops. The second term (i.e., $M \cdot W_S$) does not utilize any spectral resources because it represents the gateway's own access traffic transmitted directly over cables to the core.

It should be noted that in a UDN, the number of small cells is usually too large. As the traffic is forwarded to a gateway, each hop utilizes a wireless channel which carries more and more traffic as it nears the gateway. The total required spectrum resources increase as the number of hops increases. Each bit is transmitted many times, once for each hop, and this redundancy wastes spectral resources. Thus, minimizing the number of hops is essential to increasing the spectral utilization efficiency and thereby the capacity of the backhaul network to carry more information bits per unit time. Minimizing the ANH is achieved by optimizing the gateway locations.

### V. MATHEMATICAL FORMULATION
The gateway location problem in the distributed wireless backhaul for 5G ultra-dense small cell networks is similar to the $p$-median problem of selecting $p$ centers or medians and allocating all other points to their nearest centers with the goal of minimizing the sums of the distances between the centers and their assigned points. Each center is chosen from among the given points.

The complexity of this GLP problem is $O(N^3+(N^2 log^{p+2}))$, i.e., the complexity of calculating shortest path trees for all nodes plus the complexity of the *p*-median problem. This shows that as the size of the problem instance increases, it rapidly becomes too large to solve.

Mathematically, the GLP problem for 5G ultra-dense small cell networks can be summarized as follows:

**Inputs:**

$d_{ij}$ = Distance between small cell *i* and gateway *j*, in number of hops
*p* = Number of gateways to locate

**Decision Variables:**

$$X_j = \begin{cases} 1, & \text{if we locate gateway at small cell } j \\ 0, & \text{if not} \end{cases} \quad (3)$$

$$Y_{ij} = \begin{cases} 1, & \text{if small cell } i \text{ is served by gateway } j \\ 0, & \text{if not} \end{cases} \quad (4)$$

**Minimize:**

$$\sum_i \sum_j d_{ij} Y_{ij} \quad (5)$$

**subject to:**

$$\sum_j Y_{ij} = 1 \quad \forall i \in N \quad (6)$$

$$\sum_j X_j = p \quad (7)$$

$$Y_{ij} - X_j \leq 0 \quad \forall j \in M; i \in N \quad (8)$$

- (3) and (4) specify that the decision variables (location variables $X_j$ and allocation variables $Y_{ij}$) are binary.
- The objective (5) minimizes the total weighted distance (hops) of all small cells to the assigned gateways.
- Constraint (6) ensures that each small cell is assigned to exactly one gateway.
- Constraint (7) ensures that exactly *p* gateways are selected.
- Constraint (8) ensures that any small cell *i* is assigned only to a location that is a gateway ($X_j$ = 1).

This integer linear programming problem is solved by the CPLEX solver [76]. The input data are:
- A finite number of small cells with locations.
- A finite number of possible gateway locations. In our case any small cell can be selected as a gateway.
- A distance matrix $d[N][N]$ of dimension $N \times N$, where $N$ is the number of small cells including potential gateways. Each element $d_{ij}$ represents the smallest number of hops between small cell *i* and small cell *j*, calculated via Dijkstra's algorithm where each potential link has nominal length 1. A shortest route tree formulation reduces the calculation effort.

Solving this problem provides the minimum total distance (in hops) from gateways to small cells along with the optimal gateway locations. The minimum ANH is calculated using:

$$ANH = Total\ minimum\ distance\ /\ (N - M) \quad (9)$$

## VI. PROPOSED K-GA ALGORITHM

Optimal algorithms can solve only small instances of the *p*-median problem in reasonable time, so heuristic solutions are needed. Complete bibliographies for several meta-heuristic techniques for the *p*-median problem are given in [16][77]. Important meta-heuristics methods to solve the *p*-median problem are: (i) Tabu Search, (ii) Variable Neighborhood Search, (iii) Genetic Algorithms, (iv) Scatter Search, (v) Simulated Annealing, (vi) Heuristic Concentration, (vii) Ant Colony Optimization, and (viii) Neural Networks. Genetic Algorithms are most widely used to solve the *p*-median problem [78][79][80][81][82][83]. The GA initial population is chosen randomly, but this gives a higher likelihood of being trapped at a local optimum. A better-chosen initial population can provide higher quality chromosomes, leading to better final solutions.

There has been very little research on methods that generate a better initial population for GA to solve the *p*-median problem, especially in the case of 5G ultra-dense networks. Generating a better initial population for GA using *K*-means clustering is an interesting research area [84][85]. Motivated by the effectiveness of GAs for the *p*-median problem and the efficiency of *K*-means, a novel heuristic K-GA is proposed here to solve the *p*-median problem in the context of the gateway location problem in 5G ultra-dense networks. The K-GA algorithm has three phases. The steps are listed in Algorithm 1.

### A. Phase 1: K-means Clustering

We start with the unsupervised *K*-means ML algorithm applied to the unlabeled elements (i.e. the elements not assigned to any group or cluster). We use the squared Euclidean distance metric. The algorithm has two stages:

**Stage 1** finds the centroids given the number of clusters and the data points, and then associates each small cell with the nearest centroid.

**Stage 2** updates the centroids. New centroids are calculated by taking the average of the locations of the small cells associated with the cluster. The small cells are then reassigned to the new centroids.

The process is iterated within a replication until no further changes in centroids or small cell associations occur [86]. $R_{max}$ is the maximum number of replications and controls the number of times the clustering process is repeated. For each replication, the initial centroids are selected randomly. The best result from all replications is selected as the final result of this stage. It consists of *M* centroids that have the smallest sum, over all clusters, of the within-cluster sums of small-cells-to-cluster-centroid distances.

### B. Phase 2: Combining K-means and GA

In this phase, we generate an initial population for the next phase of K-GA. Using the *K*-means result, we select the $t_1$,

---

**Algorithm 1. K-GA Heuristic**

**Inputs:**
- Locations of small cells
- $M$: number of gateways to be chosen
- $t$: number of small cells closest to each $K$-means centroid to be selected
- $R_{max}$: maximum number of $K$-means replications
- $P_m$: GA mutation probability
- $G_{max}$: GA maximum number of generations

**Output:**
- Gateway locations

**BEGIN**
1. **Do** $R_{max}$ times:
2.    Arbitrarily choose $M$ initial centroids
3.    Repeat until convergence is achieved:
4.       Assign each small cell to the cluster having the closest centroid
5.       Calculate the new centroid of each cluster
6.    Save the result of this replication
7. Find the best result among all replications and extract locations of the $M$ centroids
8. Select $t$ small cells nearest to each of the $M$ centroids
9. Generate $t^M$ small cell combinations as the initial population of chromosomes
10. Calculate *fitness* of each chromosome in the initial population
11. **Do** $G_{max}$ times:
12.    Perform *selection* process $t^M$ times to generate the mating pool
13.    **Do** $t^M/2$ times:
14.       Perform *crossover* operation to generate two new chromosomes
15.       Perform *mutation* operation using $P_m$ on each new chromosome
16.       Perform *repair* operation as necessary
17.       Calculate *fitness* of the two new chromosomes
18.    Find and save the chromosome with the best fitness in this generation
19.    Replace current population with the new population
20. Extract and output the gateway locations from the saved best chromosome
**END**

---

$t_2, t_3, ..., t_M$ small cells nearest to the $M$ centroids. Note that the number of small cells associated with each centroid could differ. In this work, $M = 4$ and $t_1 = t_2 = t_3 = t_4 = 4$.

### C. Phase 3: Genetic Algorithm

A Genetic Algorithm [87] comprises the last stage of K-GA. Since there are $N$ small cells, an individual GA *chromosome* is a binary string of length $N$ with one position for each small cell. A '1' indicates that the associated small cell is a gateway, and a '0' indicates that it is not. GAs require an initial *population* of different chromosomes, each encoding a solution specifying the location of the $M$ gateways.

The GA population size is determined by the process of generating the initial population, as follows. Phase 1 returns $M$ clusters with $M$ centroids, and Phase 2 returns the sets of $t$ small cells nearest to each centroid. We generate an initial population of $t^M$ chromosomes by listing all combinations of one gateway taken from each of the Phase 2 sets. Using $M = t = 4$ gives $4^4 = 256$ combinations for gateway locations, which constitutes the initial population of 256 chromosomes.

The *fitness* value of a chromosome is the ANH associated with the solution it encodes. The fitness function runs the shortest path algorithm to assign small cells to gateways and generate shortest path trees for calculating ANH as the fitness value. A small cell closest to a GW in terms of number of hops becomes part of the shortest path tree rooted at that GW. Lower ANH indicates better fitness.

We use roulette wheel selection, single-point crossover and 1% mutation probability. A *repair procedure* may be needed after crossover and mutation if the number of gateways in a new chromosome is not equal to $M$. If there are too many gateways, then randomly chosen '1s' equal to the number of extra GWs are converted to '0'. If there are insufficient gateways, then randomly chosen '0s' equal to the number of missing GWs are converted to '1'.

After mutation, the fitness values are calculated. The process of selection, crossover and mutation repeats until enough new individuals have been produced to create a new generation.

The selection, crossover, and mutation processes continue for $G_{max}$ generations. The chromosome having the best fitness over all generations is output, providing the final gateway locations and allowing the calculation of ANH and BNC using (2) and (1), respectively.

*a: K-GA COMPUTATIONAL COMPLEXITY*
K-GA's complexity is $O(NI + Gn_{pop}N^2/M)$. It takes $O(NI)$ time to generate the best cluster centers for the initial GA population, where $N$ is the number of small cells and $I$ is the maximum number of iterations for the $K$-means algorithm. The complexity of the final stage of K-GA is $O(Gn_{pop}N^2/M)$, where $G$ is the number of generations for GA, $n_{pop}$ is the number of chromosomes, and $M$ is the number of gateways. There are on average $N/M$ small cells associated with each of the $M$ gateways, each requiring a Dijkstra's shortest route tree solution of complexity $O(N/M)^2$. Generally, $M << N$ and the complexity of K-GA reduces to $O(NI + N^2Gn_{pop})$.

## VII. METHODOLOGY

### A. Network Topology
We consider a UDN in a circular area with a radius of 1,000 meters for all three distribution scenarios. Small cells are placed using a homogenous Poisson Point Process (PPP) scheme [88]. The number of small cells is a Poisson random variable with mean $\lambda$, where $\lambda$ represents the average number of small cells in the circular area. To simulate different access traffic distributions, the network topology is simulated in three different scenarios: Uniform Distribution (UD), bivariate Gaussian Distribution (GD), and Cluster Distribution (CD).

The transmission range for all small cells is considered as 200 meters in all three distribution setups [9][65]. Millimeter wave backhaul does not allow small cells to have larger transmission ranges because of high path loss. Generally, a 200 m coverage radius is considered ideal for small cells [5][89][90]. Small cells are placed apart from each other to avoid complete overlap of their backhaul coverage areas. While generating a network distribution, we enforce restrictions on network topologies to maintain a certain minimum inter-small cell distance. We consider different minimum inter-small cell distances in different network distributions to represent various small cell distribution patterns.

#### 1) UNIFORM DISTRIBUTION SCENARIO
In the circular area, the locations of small cells are generated using a uniform random distribution, i.e., nodes are distributed uniformly and randomly in all directions. The uniform distribution is widely used because of its simplicity and systematic flexibility. An example of a topology created using a uniform random distribution is shown in Fig.2. Small cells are separated by a minimum distance of 50 m to avoid complete overlap of coverage area between adjacent cells.

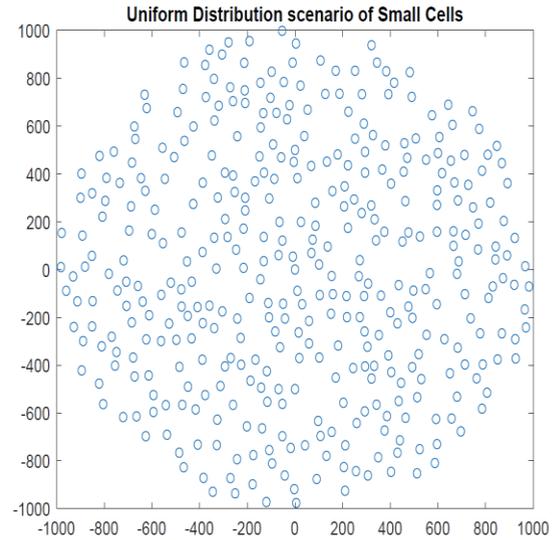

FIGURE 2. UD Network Topology

#### 2) BIVARIATE GAUSSIAN DISTRIBUTION SCENARIO
Small cells are distributed in the circular region according to a symmetric bivariate 2D Gaussian with a peak at the area center. This makes the simulation environment more realistic.

For example, it is a good model of a city where the downtown (center of the city) has more users compared to the outskirts of the city. The bivariate Gaussian distribution is completely determined by its parameters ($\mu$ and $\sigma^2$), the expected value and variance for the random variable [75].

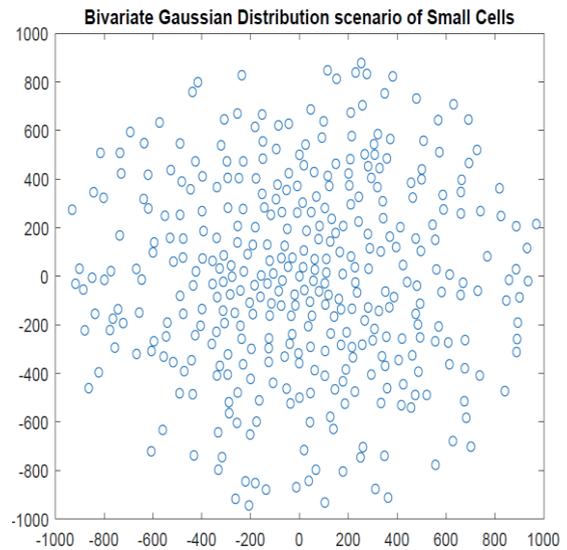

FIGURE 3. GD Network Topology

We set $\mu = 0$ and the standard deviation is normalized to 0.40 to provide a narrower and denser distribution around the center. The theoretical distribution of a Gaussian extends to infinity, but we limit the area to a radius of 1,000 m. If the radius is large enough compared to the standard deviation, the edge effect is negligible (the probability of generating a point outside the service area is very small). A topology generated using a bivariate Gaussian distribution is shown in Fig.3. Small cells are separated by a minimum distance of 40 m.

### 3) CLUSTER DISTRIBUTION SCENARIO

In this scenario, more nodes are generated in specific regions, which represents more users gathered at 'hotspots', which may be offices, universities, shopping malls, etc. The cluster distribution to model this topology is a combination of groups of small cells (clusters) and uniformly distributed small cells. This provides a scenario closer to real world situations where more users are clustered at particular places, and others are also present outside these regions. Small cells are usually deployed based on traffic demands per unit area, where more users clustered together indicate higher traffic and denser small cell deployment.

A cluster distribution topology is constructed in two phases. First, we generate clusters within the original circular area. To get the first cluster's center coordinates, a random angle $\theta$ on a 500 m radius circle is generated and the coordinates of the center are derived as

$$x = radius * cos(\theta), \quad (10)$$

$$y = radius * sin(\theta). \quad (11)$$

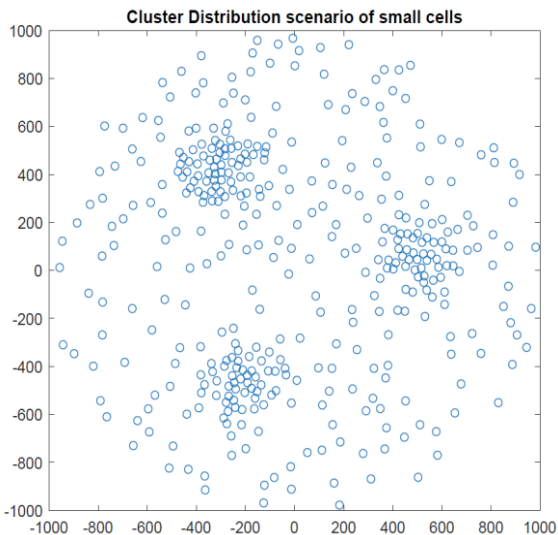

**FIGURE 4.** CD Network Topology with 3 Clusters

We generate locations of small cells within a 100 m radius around the first cluster center using a bivariate random Gaussian distribution with a peak at the cluster center and standard normal parameters $\mu = 0$ and $\sigma^2 = 1$. Small cells in a cluster are separated by a minimum distance of 25 m. For the center of the next cluster, we derive a new $\theta$ as follows:

$$\theta_{new} = \theta + \frac{360°}{\text{Number of clusters needed in a topology}} \quad (12)$$

Using equations (10) and (11), new cluster center coordinates are generated as described earlier. This process is repeated for the required number of clusters. The number of nodes in each cluster is different. In the second phase of the CD procedure, nodes are generated using a uniform random distribution, with a minimum distance of 50 m separating the small cells. A cluster distribution topology with 3 clusters is shown in Fig.4. In this paper, we generate CD topologies with 6 clusters at all node densities.

### B. Connectivity Graph

We create the neighbor table for each small cell based on the data received from the *Hello* messages broadcast by its neighbors [91]. Each "Hello message" contains a small cell's ID and coordinates. A small cell lists other small cells in its transmission range in its neighbor table once it receives their "Hello messages". Using the neighbor tables of all small cells, we generate the connectivity graph and ensure that the resulting network is fully connected. If there is a least one isolated node in the network, that topology is discarded and replaced.

### C. Shortest Path Trees

We employ *Dijkstra's shortest path algorithm* to find the minimum number of hops from a small cell to its assigned gateway. The locations of GWs (available at the output of an algorithm) and the connectivity graph are input to Dijkstra's algorithm, and shortest path trees are found from gateways to small cells. Setting all small cell connection "distances" to 1 in the connectivity graph means that route lengths are measured in number of hops. Finally, we calculate the number of hops from all small cells to their GWs in these trees and take their average as ANH.

### D. Comparators

We implemented several methods for comparison with K-GA: a genetic algorithm, *K*-means, *K*-medoids, a baseline approach, and the combination of *K*-medoids and a genetic algorithm (*KM-GA*). We also implemented an optimal ILP to find the exact GLP solution.

#### 1) GENETIC ALGORITHM

Its implementation is identical to the GA stage in the K-GA algorithm except for the generation of the initial population, which is done by randomly generating *M* 1s in each chromosome.

#### 2) *K*-MEANS ALGORITHM

Based on the final output of *M* centroids of *M* clusters from the *K*-means algorithm, we choose the small cell nearest to each centroid as a Gateway. This set of *M* GW locations is used as input to the shortest path algorithm for calculating the ANH.

### 3) K-MEDOIDS ALGORITHM
The best result from all replications is selected as the result of the *K*-medoid clustering algorithm. It consists of *M* medoids that have the smallest sum, over all clusters, of the within-cluster sums of small cells-to-cluster medoid distances. The final *M* medoids (or small cells) are selected as gateway locations for calculating ANH.

### 4) BASELINE METHOD
Gateways are placed at fixed locations at equal distances around a 500 m radius circle within the original circular area for the uniform random distribution and bivariate Gaussian distributions. For the cluster distribution, the small cells nearest to the cluster centers are chosen as GWs. From among *C* nearest small cells of cluster centers, *M* gateways are chosen, where *C* is the total number of clusters present in a topology and $C \geq M$. These gateway locations are provided as input to Dijkstra's shortest path algorithm to obtain ANH.

### 5) COMBINED K-MEDOIDS AND GENETIC ALGORITHM (KM-GA)
KM-GA is identical to K-GA except for the 1st phase, where an actual small cell is returned as a cluster center (or medoid). Using the *K*-medoids output, we select the *M* medoids and $t_1-1, t_2-1, t_3-1, …, t_M-1$ small cells nearest to the *M* medoids to generate the initial population for the GA. The complexity of KM-GA is $O(N^2I + NGn_{pop})$, where *N* is the total number of small cells, *G* is the number of generations for GA, and npop is the number of chromosomes. *I* is the number of replications for *K*-medoids.

Based on the computational complexity and runtime results of both algorithms, K-GA proves better than KM-GA.

### 6) OPTIMAL ILP
The exact ILP solution is provided by the CPLEX mixed-integer solver [76] using a model formulated in the OPL Optimization Programming Language [92]. OPL is part of the CPLEX software package.

## VIII. PERFORMANCE EVALUATION

### A. Simulation Environment
The *p*-median problem is formulated and solved in IBM ILOG CPLEX Optimization Studio [76]. Extensive simulations for the heuristic approaches (K-GA, GA, K-medoids, K-means, KM-GA and Baseline) are implemented and executed using MATLAB R2019b [93]. We calculate the backhaul network capacity using (1) where $W_S$ is 1 Gbps and $W_G$ is 100 Gbps [65][94].

We test 5 different node densities of small cells in a circular area having a radius of 1,000 meters for uniform distribution, bivariate Gaussian distribution and cluster distribution. For convenience we define node density as the population of small cells in the 1,000 meters radius test area. To realize the ultra-dense deployment scenario, a higher small cell density per 1 km radius macrocell coverage area is used, as is highly anticipated [5][9][65]. We consider node densities from 310 to 470 in this paper. A study shows that 40% of the operators are expected to deploy nearly 350 small cells per square kilometer [95], while another study shows that expected numbers for small cells deployment are mostly between 251–500 in 2020–2025 for different scenarios [96]. For each node density in all distribution scenarios, we generate 100 different topologies as a part of our Monte Carlo simulation setup. Overall, to evaluate all scenarios, we generate 1,500 different network topologies and solve via all methods separately on each topology. We calculate the mean value of the ANH and the BNC for the 100 topologies at each node density in all distribution scenarios and compute the 95% confidence intervals (CIs) as well. We use 4 gateways for all node densities.

Parameter settings are as follows:
- *K-means*: 100 replications for each node density.
- *K-medoids*: 100 replications for each node density.
- *GA*: 100 generations for each node density. The population size is 300 for all node densities (310 to 470).
- *K-GA*: 50 replications of *K*-means and 50 GA generations for each node density. Population size 256 is used for all node densities because a constant size initial population is generated from the *K*-means stage.
- *KM-GA:* 50 replications of *K*-medoids and 50 GA generations for each node density. Population size 256 is used.
- *Baseline*: On the 500 meter circle for UD and GD, the 4 gateway locations are generated at the following coordinates: (294, 405), (-294, 405), (-294, -405), (294, -405). For CD, gateways are taken as the nearest small cells of the center of the cluster 2, 3, 5, and 6. These gateway locations remain the same for all topologies for all node densities.

### B. Simulation Results
#### 1) RESULTS FOR UNIFORM DISTRIBUTION
We also implemented the combination of *K*-medoid algorithm with a genetic algorithm to confirm the effectiveness of K-GA. In this section we present two different analyses in uniform distribution scenarios. First, we compare K-GA with KM-GA and subsequently we compare K-GA algorithm with the five other methods.

##### a: COMPARISON OF K-GA AND KM-GA
To compare K-GA with KM-GA, we consider node densities of 310, 350, 390, 430 and 470. We evaluate and compare the results for both algorithms based on their mean values of ANH and runtimes. FIGURE 5 and Table 2 show the ANH and runtimes for both methods. K-GA performs consistently better compared to KM-GA for all node densities but the difference in performance is small. K-GA chooses the small cells nearest to *M* centroids to generate the initial population for GA. This provides more diverse locations for the initial GA population as compared to using medoids and small cells closer to medoids to generate the initial population in KM-GA. The more diverse initial K-GA population helps the GA perform better. K-GA is also faster than KM-GA. They perform closely in terms of ANH, but K-GA is better than KM-GA in terms of computational complexity and runtime.

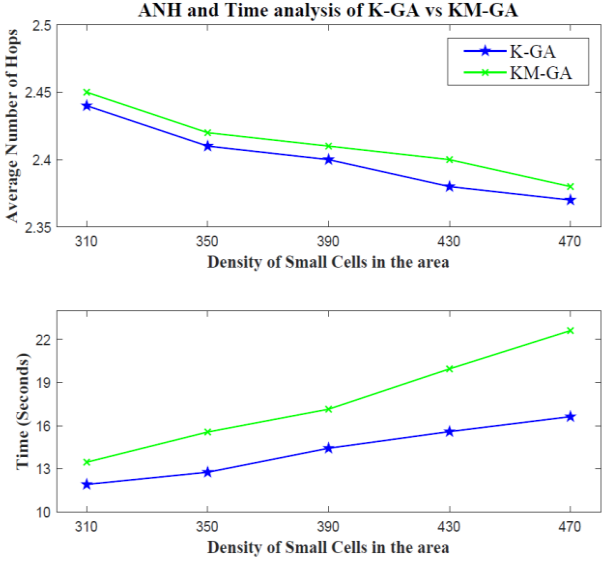

**FIGURE 5.** ANH and Time Analysis for K-GA and KM-GA

### b: COMPARISON OF K-GA WITH 5 OTHER METHODS

This section compares K-GA with five other methods in terms of ANH and BNC. We simulate 5G ultra-dense networks according to our network model with 4 gateways and small cell densities ranging from 310 to 470. As an example, the gateway locations selected by all methods for a single topology are plotted in Fig.6. The *K*-means and *K*-medoids differ in a single gateway location, but the other solutions differ more significantly. No method finds all 4 optimum locations.

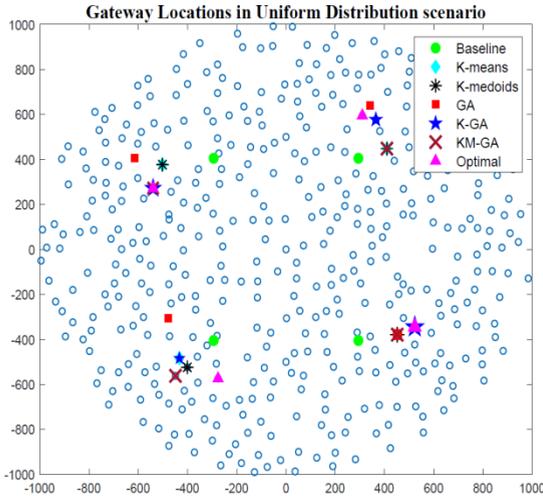

**FIGURE 6.** Example GW Locations in UD Scenario

Fig.7 plots ANH vs. small cell density. As the node density increases, connectivity increases, enabling the shortest path algorithm to find better paths resulting in smaller ANH at higher node densities. ANH for K-GA is significantly smaller than the baseline method. At a lower node density of 310, ANH ranges from 3.24 (Baseline) to 2.51 (*K*-means) to 2.50 (GA) to 2.49 (*K*-medoids) to 2.44 (K-GA). *K*-medoids works slightly better than *K*-means because of its more effective clustering and gateway selection mechanism. GA performs well compared to both *K*-means and *K*-medoids at lower node densities. In *K*-means, gateway locations are based on *M* centroids, so at lower node densities, hops are longer due to larger distances from centroids to small cells. This reduces for *K*-medoids and GA. In contrast, *K*-medoids returns actual small cells as GWs and GA works directly with the number of hops as the fitness measure. In general, GA, *K*-means and *K*-medoid's CIs are overlapping, and they perform similarly. K-GA improves over GA due to the better initial population provided by its *K*-means stage. K-GA improves performance by *24.69%*, *2.79%*, *2.4%,* and *2%* compared to baseline, *K*-means, GA, and *K*-medoids respectively.

At a higher node density of 470, ANH ranges from 2.53 (Baseline) to 2.42 (GA) to 2.41 (*K*-means and *K*-medoids) to 2.37 (K-GA). This is an increase in performance for K-GA of *6.32%, 1.66%, 1.66%,* and *2.07%* compared to baseline, *K*-means, *K*-medoids, and GA respectively.

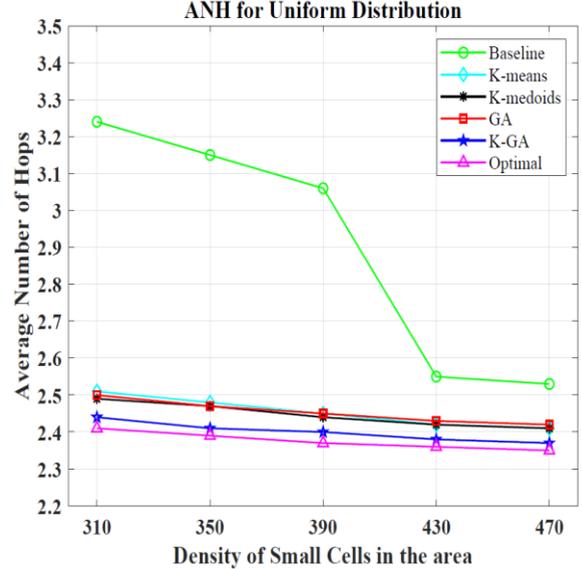

**FIGURE 7.** Average Number of Hops in UD Scenario

At higher node densities, *K*-means and *K*-medoids perform similarly to GA and baseline also shows significant improvement. This is because the smaller distances from gateways to small cells provide more connectivity options. K-GA outperforms the other methods in terms of ANH at all node densities.

We also evaluate the quality of K-GA by comparison with the exact solution obtained from the ILP. As shown in

TABLE 2
ANH AND RUNTIME ANALYSIS OF K-GA VERSUS KM-GA

| Small Cell Density | K-GA | | | | KM-GA | | | |
|---|---|---|---|---|---|---|---|---|
| | ANH | | Runtime (seconds) | | ANH | | Runtime (seconds) | |
| | Mean | 95% CI | Mean | 95% CI | Mean | 95% CI | Mean | 95% CI |
| 310 | 2.44 | 2.43—2.45 | 11.91 | 11.73—12.10 | 2.45 | 2.44—2.46 | 13.46 | 13.30—13.61 |
| 350 | 2.41 | 2.40—2.42 | 12.76 | 12.65—12.87 | 2.42 | 2.41—2.43 | 15.56 | 14.81—16.31 |
| 390 | 2.40 | 2.39—2.40 | 14.43 | 14.28—14.59 | 2.41 | 2.40—2.41 | 17.15 | 16.98—18.33 |
| 430 | 2.38 | 2.38—2.39 | 15.59 | 15.44—15.75 | 2.40 | 2.40—2.41 | 19.96 | 19.65—20.26 |
| 470 | 2.37 | 2.37—2.38 | 16.63 | 16.50—16.75 | 2.38 | 2.37—2.38 | 22.62 | 22.46—23.26 |

TABLE 3
ANH FOR BASELINE, *K*-MEDOIDS, *K*-MEANS AND GA IN UD SCENARIO

| | Average Number of Hops | | | | | | | |
|---|---|---|---|---|---|---|---|---|
| Small Cell Density | GA | | *K*-means | | *K*-medoids | | Baseline | |
| | Mean | 95% CI | Mean | 95% CI | Mean | 95% CI | Mean | 95% CI |
| 310 | 2.50 | 2.49—2.50 | 2.51 | 2.49—2.54 | 2.49 | 2.48—2.51 | 3.24 | 3.17—3.32 |
| 350 | 2.47 | 2.46—2.48 | 2.48 | 2.46—2.49 | 2.47 | 2.45—2.48 | 3.15 | 3.09—3.22 |
| 390 | 2.45 | 2.45—2.46 | 2.45 | 2.44—2.47 | 2.44 | 2.43—2.45 | 3.06 | 2.99—3.12 |
| 430 | 2.43 | 2.43—2.44 | 2.42 | 2.42—2.43 | 2.42 | 2.42—2.43 | 2.55 | 2.54—2.56 |
| 470 | 2.42 | 2.42—2.43 | 2.41 | 2.41—2.42 | 2.41 | 2.41—2.42 | 2.53 | 2.53—2.54 |

TABLE 4
ANH FOR K-GA AND OPTIMAL APPROACH IN UD SCENARIO

| | Average Number of Hops | | | | |
|---|---|---|---|---|---|
| Small Cell Density | Optimal | | K-GA | | Gap between Optimal and K-GA (%) |
| | Mean | 95% CI | Mean | 95% CI | |
| 310 | 2.41 | 2.40—2.42 | 2.44 | 2.43—2.45 | 1.24% |
| 350 | 2.39 | 2.38—2.40 | 2.41 | 2.40—2.42 | 0.84% |
| 390 | 2.37 | 2.35—2.38 | 2.40 | 2.39—2.40 | 1.27% |
| 430 | 2.36 | 2.36—2.37 | 2.38 | 2.38—2.39 | 0.85% |
| 470 | 2.35 | 2.35—2.36 | 2.37 | 2.37—2.38 | 0.85% |

TABLE 5
RUNTIME ANALYSIS FOR K-GA AND OPTIMAL APPROACH IN UD SCENARIO

| | Runtime (seconds) | | | | |
|---|---|---|---|---|---|
| Small Cell Density | Optimal | | K-GA | | Runtime saving with K-GA (%) |
| | Mean | 95% CI | Mean | 95% CI | |
| 310 | 228.34 | 227.99—228.69 | 11.91 | 11.73—12.10 | 94.78% |
| 350 | 242.71 | 242.41—243.01 | 12.76 | 12.65—12.87 | 94.74% |
| 390 | 262.54 | 261.89—263.19 | 14.43 | 14.28—14.59 | 94.50% |
| 430 | 277.61 | 275.62—279.61 | 15.59 | 15.44—15.75 | 94.38% |
| 470 | 326.68 | 322.32—331.04 | 16.63 | 16.50—16.75 | 94.91% |

Table 4, K-GA performs within *1.24%* of optimal ANH at low node density (310) and within *0.85%* for high node density (470). We calculate the average runtime for each node density for K-GA and the optimal solution in the uniform distribution scenario, as shown in Table 5 and Fig.8. The ILP requires a very large amount of runtime. In contrast, K-GA finds near-optimal solutions (within *2%*) very quickly in almost *95%* less time. Mean values and CI for ANH for baseline, *K*-means, *K*-medoids and GA are given in Table 3, while results for the ILP and K-GA are given in Table 4 for the uniform distribution scenario.

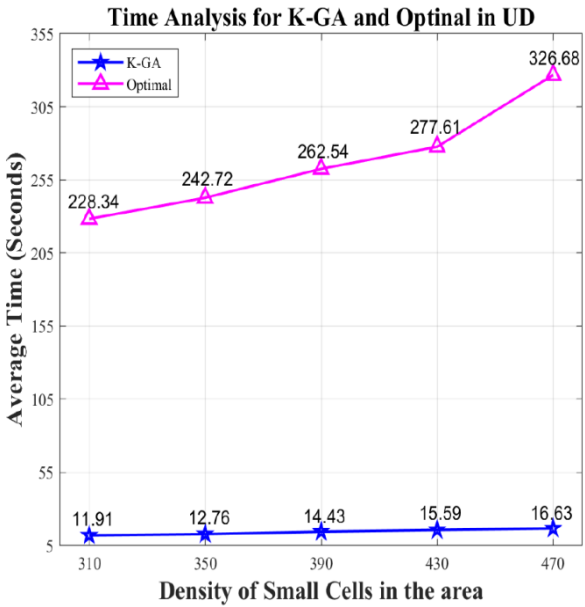

**FIGURE 8.** Time Analysis of K-GA and ILP in UD Scenario

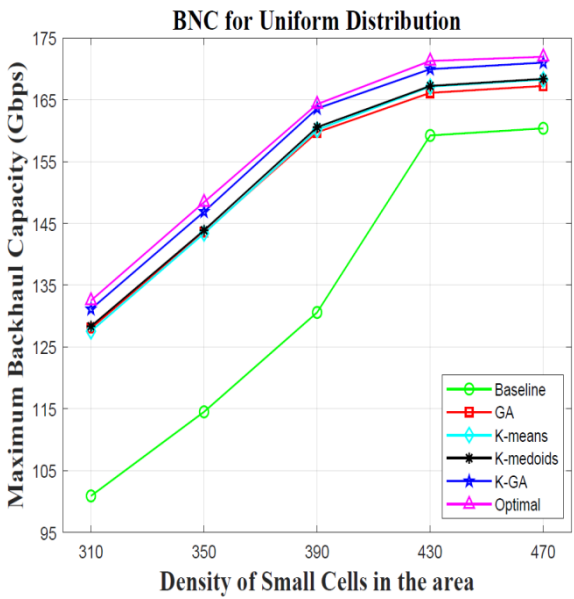

**FIGURE 9.** Backhaul Network Capacity in UD Scenario

Fig.9, Table 6, and Table 7 illustrate the BNC for all methods. With increasing node density, ANH decreases, resulting in increased BNC for all methods. At a lower node density of 310, BNC is 100.88 (Baseline), 127.53 (*K*-means), 128.01 (GA), 128.34 (*K*-medoids), and 131.11 (K-GA). This indicates a capacity improvement of *29.97%, 2.81%, 2.42%,* and *2.16%* for K-GA compared to baseline, *K*-means, GA, and *K*-medoids respectively. At a higher node density of 470, K-GA shows an improvement in capacity of *6.63%, 1.63%,*
*1.56%,* and *2.26%* compared to baseline, *K*-means, *K*-medoids, and GA respectively. K-GA obtains backhaul network capacity within *2%* of the optimal solution for all node densities in the uniform distribution scenario. At the high node density of 470, K-GA achieves BNC within *0.55%* of the optimal solution.

### 2) RESULTS FOR BIVARIATE GAUSSIAN DISTRIBUTION

This section summarizes the results for the bivariate Gaussian distribution scenario. High node densities of small cells are considered (from 310 to 470 with an interval of 40). To illustrate gateway locations found in this scenario, results for all methods for a single topology are plotted in Fig.10. *K*-means and *K*-medoids have 2 identical GW locations and GA has totally different locations compared to them. K-GA shares 3 gateway locations with the optimal solution. K-GA generally obtains similar results as the optimum ILP in the Gaussian distribution scenario.

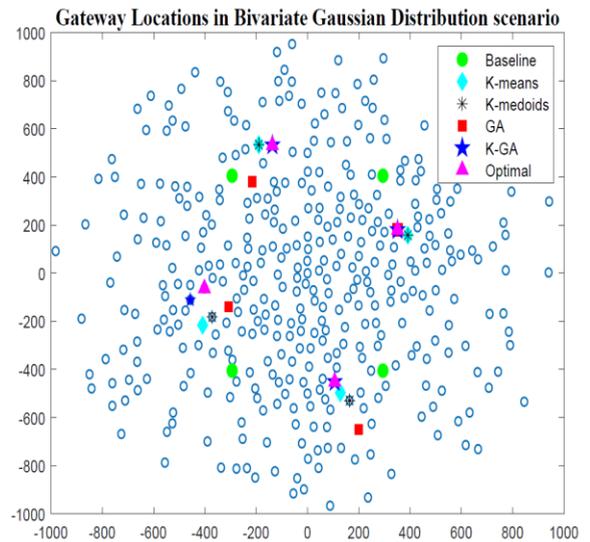

**FIGURE 10.** Example GW Locations in GD Scenario

In Fig.11, the average number of hops is plotted for all heuristic approaches, baseline method and optimal solution in the bivariate Gaussian distribution scenario. When the network size increases from 310 nodes to 470 nodes, the difference between the ANH for baseline, *K*-means, *K*-medoids, GA, and K-GA is about 0.01–0.05, which is less than in the uniform distribution scenario.

TABLE 6
BNC FOR BASELINE, *K*-MEDOIDS, *K*-MEANS AND GA IN UD SCENARIO

| Backhaul Network Capacity (Gbps) | | | | | | | | |
|---|---|---|---|---|---|---|---|---|
| Small Cell Density | GA | | *K*-means | | *K*-medoids | | Baseline | |
| | Mean | 95% CI | Mean | 95% CI | Mean | 95% CI | Mean | 95% CI |
| 310 | 128.01 | 126.50—129.53 | 127.53 | 125.63—129.43 | 128.34 | 126.67—130.02 | 100.88 | 98.23—103.53 |
| 350 | 143.54 | 141.80—145.28 | 143.36 | 141.37—145.36 | 143.87 | 141.86—145.88 | 114.49 | 111.88—117.11 |
| 390 | 159.73 | 158.42—161.04 | 160.14 | 158.46—161.81 | 160.55 | 159.12—161.98 | 130.58 | 127.70—133.45 |
| 430 | 166.12 | 165.64—166.60 | 167.12 | 166.56—167.68 | 167.23 | 166.66—167.80 | 159.22 | 158.57—159.87 |
| 470 | 167.23 | 166.87—167.58 | 168.26 | 167.89—168.64 | 168.38 | 167.99—168.77 | 160.37 | 159.93—160.80 |

TABLE 7
BNC FOR K-GA AND OPTIMAL APPROACH IN UD SCENARIO

| Backhaul Network Capacity (Gbps) | | | | | |
|---|---|---|---|---|---|
| Small Cell Density | Optimal | | K-GA | | Gap between Optimal and K-GA (%) |
| | Mean | 95% CI | Mean | 95% CI | |
| 310 | 132.5 | 130.99—134.07 | 131.11 | 129.56—132.66 | 1.05% |
| 350 | 148.42 | 146.60—150.23 | 146.87 | 145.02—148.71 | 1.04% |
| 390 | 164.28 | 162.23—166.32 | 163.58 | 162.23—164.93 | 0.43% |
| 430 | 171.28 | 170.81—171.76 | 169.95 | 169.42—170.48 | 0.78% |
| 470 | 171.95 | 171.67—172.23 | 171.01 | 170.71—171.31 | 0.55% |

TABLE 8
ANH FOR BASELINE, *K*-MEDOIDS, *K*-MEANS AND GA IN GD SCENARIO

| Average Number of Hops | | | | | | | | |
|---|---|---|---|---|---|---|---|---|
| Small Cell Density | GA | | *K*-means | | *K*-medoids | | Baseline | |
| | Mean | 95% CI | Mean | 95% CI | Mean | 95% CI | Mean | 95% CI |
| 310 | 2.18 | 2.17—2.19 | 2.19 | 2.18—2.20 | 2.19 | 2.18—2.20 | 2.31 | 2.29—2.32 |
| 350 | 2.18 | 2.18—2.19 | 2.19 | 2.18—2.20 | 2.19 | 2.18—2.20 | 2.29 | 2.28—2.29 |
| 390 | 2.19 | 2.18—2.19 | 2.18 | 2.18—2.19 | 2.18 | 2.18—2.19 | 2.28 | 2.27—2.28 |
| 430 | 2.19 | 2.19—2.20 | 2.18 | 2.18—2.19 | 2.19 | 2.18—2.19 | 2.26 | 2.26—2.27 |
| 470 | 2.20 | 2.20—2.21 | 2.18 | 2.18—2.19 | 2.19 | 2.18—2.21 | 2.26 | 2.25—2.27 |

TABLE 9
ANH FOR K-GA AND OPTIMAL APPROACH IN GD SCENARIO

| Average Number of Hops | | | | | |
|---|---|---|---|---|---|
| Small Cell Density | Optimal | | K-GA | | Gap between Optimal and K-GA (%) |
| | Mean | 95% CI | Mean | 95% CI | |
| 310 | 2.11 | 2.09—2.11 | 2.13 | 2.12—2.14 | 0.95% |
| 350 | 2.12 | 2.10—2.12 | 2.14 | 2.14—2.15 | 0.94% |
| 390 | 2.12 | 2.11—2.13 | 2.14 | 2.13—2.15 | 0.94% |
| 430 | 2.13 | 2.12—2.13 | 2.14 | 2.14—2.15 | 0.47% |
| 470 | 2.13 | 2.13—2.14 | 2.15 | 2.14—2.16 | 0.94% |

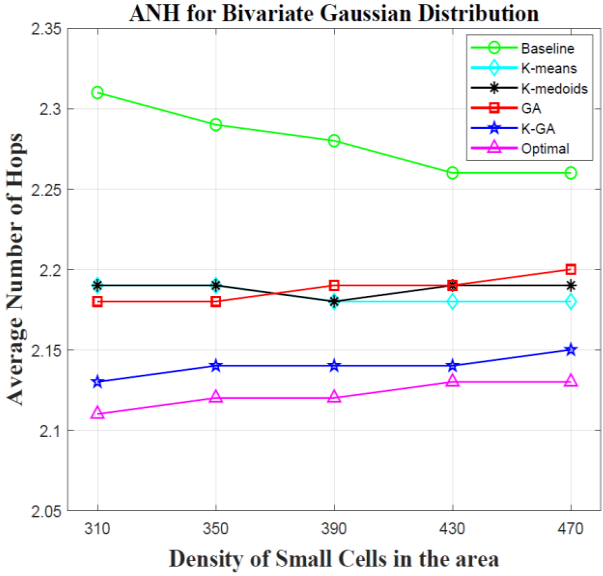

FIGURE 11. Average Numbers of Hops in GD Scenario

In the GD scenario, the probability of small cells placed near the center (and thus closer to the gateway) is high, and fewer small cells are near the border. As a result, most of the small cells need fewer hops to reach a gateway. The Baseline method has poor results compared to all other approaches. *K*-means, *K*-medoids and GA have very comparable performances. GA performs better at low node densities of 310 and 350 while *K*-means and *K*-medoids have slightly better performances at high node densities of 390, 430, and 470. K-GA provides better ANH for all node densities compared to all heuristic approaches and the baseline method. It shows an improvement of *7.79%* compared to baseline at lower node density of 310 and *4.87%* at higher node density.

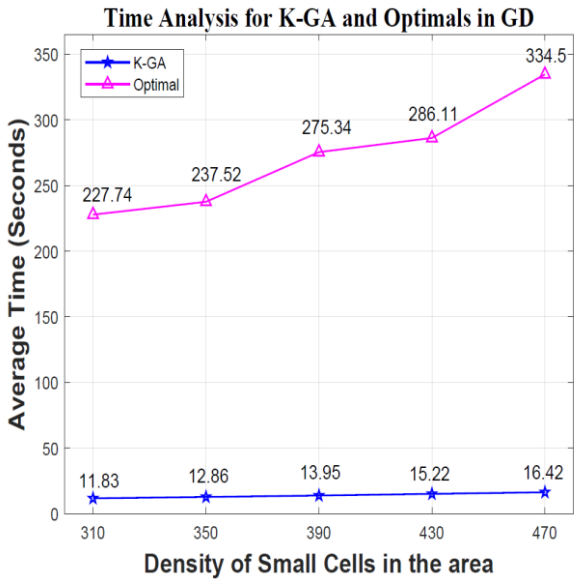

FIGURE 12. Runtimes for K-GA and Optimal ILP in GD Scenario

K-GA beats the standalone results of *K*-means and GA in ANH in this scenario. Mean values and CI for the ANH for baseline, *K*-means, *K*-medoids, and GA are given in Table 8 and ANH results for optimal ILP and K-GA are given in Table 9.

As shown in Table 9, there is an average gap of *1%* between K-GA and optimal solution for all node densities: K-GA achieves near-optimal results. Average runtimes for K-GA and the optimal ILP in GD are given in Fig.12. The optimal ILP consumes much more time than the K-GA heuristic. Overall, K-GA saves almost *95%* of runtime compared to the optimal ILP and gives results within *1%* of optimal value for each node density. Runtime analysis for K-GA and optimal in GD scenario are given in Table 10.

Fig.13 illustrates the backhaul network capacity for all approaches for the GD scenario. Compared to the baseline method, capacity improves from 138.02 to 148.68 at a node density of 310 with K-GA. At a node density of 470, capacity also improved from 179.61 to 187.76 using K-GA. Like the ANH results, *K*-means, *K*-medoids, and GA perform similarly to each other. K-GA improves over baseline, *K*-means, *K*-medoids, and GA and provides near-optimal results at all node densities. Table 11 presents the backhaul network capacity results for the comparison methods.

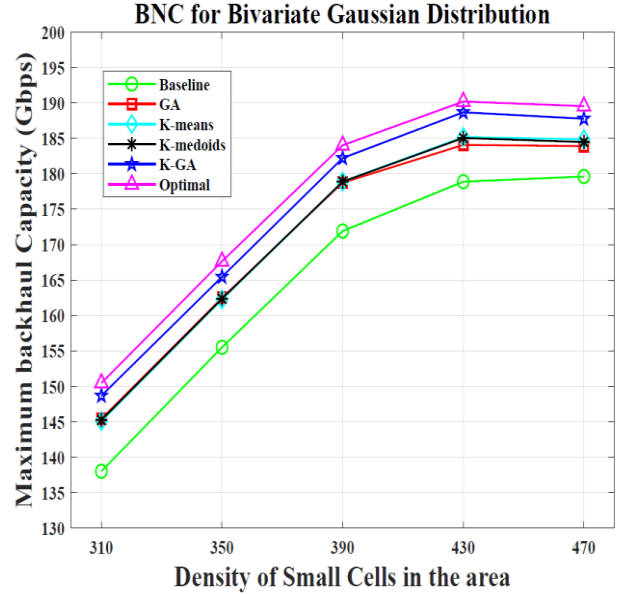

FIGURE 13. Backhaul Network Capacity in GD Scenario

As seen from Table 12, the gap between K-GA and optimal reduces from *1.19%* to *0.94%* as the node density increases. This indicates that at higher node densities K-GA performs better and provides BNCs within *1%* of optimal in the Gaussian distribution scenario.

### 3) RESULTS FOR CLUSTER DISTRIBUTION
This section compares K-GA with other approaches in the Cluster distribution scenario. As before, small cell densities range from 310 to 470 with a gap of 40. Fig.14 shows the gateway locations found by all approaches for a single 6-cluster topology. *K*-medoids, *K*-means, and GA have gateway locations that are mostly different from each other. In contrast, K-GA shares 3 gateway locations with the optimal solution. Fig.15, Table 13, and Table 14 show the average number of hops for baseline, *K*-means, *K*-medoids, genetic algorithm, K-GA, and optimal solutions in this scenario.

TABLE 10
RUNTIME ANALYSIS FOR K-GA AND OPTIMAL APPROACH IN GD SCENARIO

| Small Cell Density | Runtime (seconds) | | | | Runtime saving with K-GA (%) |
|---|---|---|---|---|---|
| | Optimal | | K-GA | | |
| | Mean | 95% CI | Mean | 95% CI | |
| 310 | 227.74 | 227.38—228.09 | 11.83 | 11.74—11.93 | 94.81% |
| 350 | 237.52 | 233.54—241.50 | 12.86 | 12.77—12.96 | 94.59% |
| 390 | 275.34 | 274.74—275.94 | 13.95 | 13.84—14.06 | 94.93% |
| 430 | 286.11 | 284.97—287.25 | 15.22 | 15.1—15.33 | 94.68% |
| 470 | 334.50 | 327.25—341.74 | 16.42 | 16.22—16.62 | 95.09% |

TABLE 11
BNC FOR BASELINE, *K*-MEDOIDS, *K*-MEANS AND GA IN GD SCENARIO

| Small Cell Density | Backhaul Network Capacity (Gbps) | | | | | | | |
|---|---|---|---|---|---|---|---|---|
| | GA | | *K*-means | | *K*-medoids | | Baseline | |
| | Mean | 95% CI | Mean | 95% CI | Mean | 95% CI | Mean | 95% CI |
| 310 | 145.42 | 143.71—147.13 | 145.10 | 143.24—146.95 | 145.25 | 143.37—147.13 | 138.02 | 136.19—139.85 |
| 350 | 162.48 | 160.93—164.02 | 162.27 | 160.62—163.92 | 162.37 | 160.74—164.01 | 155.51 | 153.88—157.15 |
| 390 | 178.74 | 177.43—180.06 | 178.89 | 177.55—180.23 | 178.88 | 177.48—180.28 | 171.91 | 170.49—173.32 |
| 430 | 184.07 | 183.60—184.54 | 185.18 | 184.61—185.76 | 185.05 | 184.47—185.63 | 178.88 | 178.35—179.41 |
| 470 | 183.90 | 183.36—184.43 | 184.86 | 184.31—185.41 | 184.47 | 183.54—185.40 | 179.61 | 179.03—180.19 |

TABLE 12
BNC FOR K-GA AND OPTIMAL APPROACH IN GD SCENARIO

| Small Cell Density | Backhaul Network Capacity (Gbps) | | | | Gap between Optimal and K-GA (%) |
|---|---|---|---|---|---|
| | Optimal | | K-GA | | |
| | Mean | 95% CI | Mean | 95% CI | |
| 310 | 150.47 | 148.74—152.20 | 148.68 | 146.88—150.47 | 1.19% |
| 350 | 167.65 | 166.0—169.21 | 165.46 | 163.88—167.03 | 1.31% |
| 390 | 184.02 | 182.65—185.39 | 182.21 | 180.84—183.59 | 0.98% |
| 430 | 190.19 | 189.74—190.64 | 188.67 | 188.18—189.17 | 0.80% |
| 470 | 189.54 | 189.06—190.02 | 187.76 | 187.22—188.30 | 0.94% |

TABLE 13
ANH FOR BASELINE, *K*-MEDOIDS, *K*-MEANS AND GA IN CD SCENARIO

| Small Cell Density | Average Number of Hops | | | | | | | |
|---|---|---|---|---|---|---|---|---|
| | GA | | *K*-means | | *K*-medoids | | Baseline | |
| | Mean | 95% CI | Mean | 95% CI | Mean | 95% CI | Mean | 95% CI |
| 310 | 2.25 | 2.24—2.26 | 2.26 | 2.24—2.27 | 2.26 | 2.25—2.27 | 2.43 | 2.41 — 2.44 |
| 350 | 2.18 | 2.18—2.19 | 2.20 | 2.19—2.21 | 2.20 | 2.19—2.21 | 2.36 | 2.34 — 2.37 |
| 390 | 2.14 | 2.13—2.14 | 2.14 | 2.13—2.15 | 2.14 | 2.13—2.15 | 2.28 | 2.26 — 2.29 |
| 430 | 2.10 | 2.09—2.10 | 2.09 | 2.08—2.10 | 2.09 | 2.08—2.10 | 2.23 | 2.21 — 2.24 |
| 470 | 2.07 | 2.06—2.08 | 2.06 | 2.05—2.07 | 2.06 | 2.05—2.07 | 2.19 | 2.18 —2.20 |

As the node density increases, ANH decreases for all approaches. Fig.14 shows that the baseline's gateway locations (small cells nearest to the cluster center) are placed near to those of other methods, although the average number of hops is much higher than for all other approaches, as shown in Fig.15. *K*-means and *K*-medoids perform similarly for all node densities in CD because of their clustering nature.

The genetic algorithm performs better at 310 and 330 node densities, but its performance degrades at higher densities (430 and 470) compared to *K*-means and *K*-medoids. At node density 390, GA, *K*-means, and *K*-medoids give the same results. The density of small cells affects their performances. Their performance appears to be similar as some of their CIs overlap. The enriched initial population provided by *K*-means leads K-GA to superior results compared to other heuristics and the baseline method. K-GA achieves near optimal results as shown in Table 14. K-GA obtains ANH within *2%* of optimal values in the cluster distribution scenario.

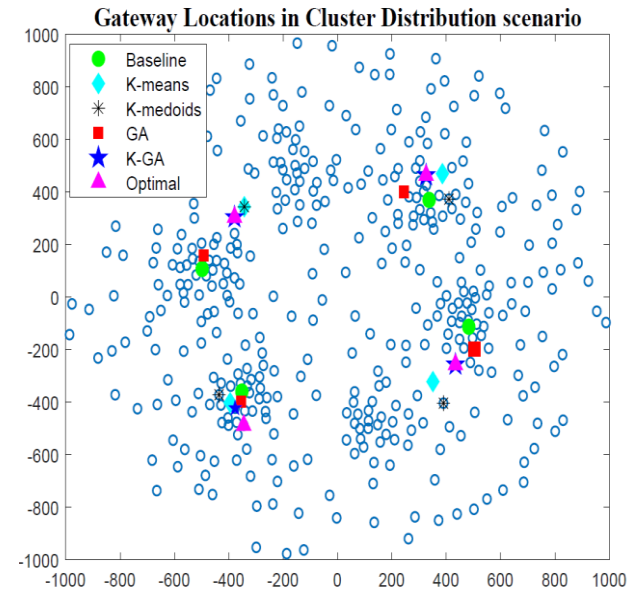

FIGURE 14. Example GW Locations in CD Scenario

The optimal solution provides slightly better results for ANH but has much longer runtimes than K-GA. K-GA saves a significant amount of runtime for larger GLPs. At a high node density of 470, K-GA takes about 16 seconds to find the solution while the optimal ILP takes 328 seconds.

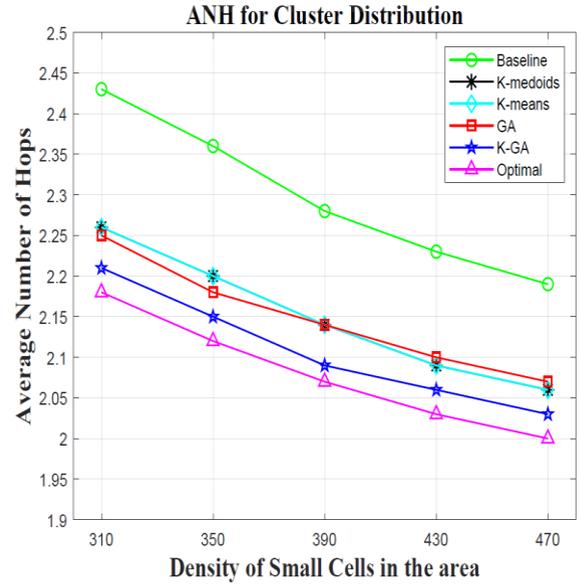

FIGURE 15. Average Numbers of Hops in CD Scenario

Table 15 gives a detailed comparison between the two solutions in terms of runtime. As shown in Table 14 and Table 15, K-GA has a maximum gap of *1.5%* in ANH while reducing runtime by more than *95%*.

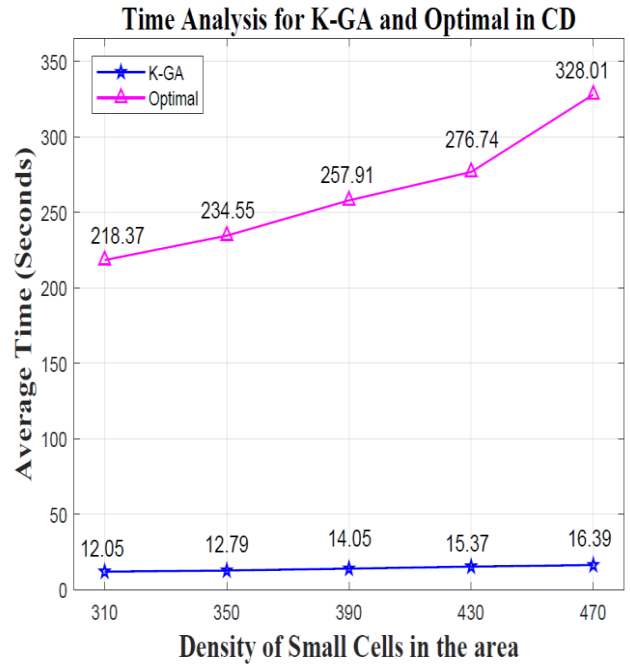

FIGURE 16. Runtimes for K-GA and Optimal ILP in CD Scenario

Fig.17, Table 16, and Table 17 show the backhaul network capacities in the cluster distribution scenario. As the node density increases, backhaul network capacity also increases. K-GA shows an improvement of *9.65%* compared to baseline at low node density of 310 and *7.77%* at high node density of 470. *K*-means, *K*-medoids, and GA perform similarly to each other at all node densities.

TABLE 14
ANH FOR K-GA AND OPTIMAL APPROACH IN CD SCENARIO

| Small Cell Density | Average Number of Hops | | | | Gap between Optimal and K-GA (%) |
|---|---|---|---|---|---|
| | Optimal | | K-GA | | |
| | Mean | 95% CI | Mean | 95% CI | |
| 310 | 2.18 | 2.17—2.19 | 2.21 | 2.19—2.22 | 1.38% |
| 350 | 2.12 | 2.11—2.13 | 2.15 | 2.14—2.16 | 1.42% |
| 390 | 2.07 | 2.06—2.07 | 2.09 | 2.09—2.10 | 0.97% |
| 430 | 2.03 | 2.02—2.04 | 2.06 | 2.05—2.07 | 1.48% |
| 470 | 2.00 | 1.99—2.01 | 2.03 | 2.02—2.04 | 1.50% |

TABLE 15
RUNTIME ANALYSIS FOR K-GA AND OPTIMAL APPROACH IN CD SCENARIO

| Small Cell Density | Runtime (seconds) | | | | Runtime saving with K-GA (%) |
|---|---|---|---|---|---|
| | Optimal | | K-GA | | |
| | Mean | 95% CI | Mean | 95% CI | |
| 310 | 218.37 | 217.76—218.98 | 12.05 | 11.84—12.26 | 94.48% |
| 350 | 234.55 | 233.69—235.42 | 12.79 | 12.67—12.92 | 94.55% |
| 390 | 257.91 | 256.82—258.99 | 14.05 | 13.89—14.21 | 94.55% |
| 430 | 276.74 | 275.48—278.0 | 15.37 | 15.22—15.51 | 94.45% |
| 470 | 328.01 | 323.69—332.33 | 16.39 | 16.21—16.58 | 95.00% |

TABLE 16
BNC FOR BASELINE, *K*-MEDOIDS, *K*-MEANS AND GA IN CD SCENARIO

| Small Cell Density | Backhaul Network Capacity (Gbps) | | | | | | | |
|---|---|---|---|---|---|---|---|---|
| | GA | | *K*-means | | *K*-medoids | | Baseline | |
| | Mean | 95% CI | Mean | 95% CI | Mean | 95% CI | Mean | 95% CI |
| 310 | 144.18 | 142.18—146.18 | 143.78 | 141.62—145.94 | 143.55 | 141.39—145.70 | 134.05 | 132.10—136.01 |
| 350 | 162.66 | 160.61—164.70 | 162.29 | 160.23—164.34 | 162.09 | 159.94—164.25 | 151.62 | 149.51—153.72 |
| 390 | 183.35 | 181.75—184.95 | 183.32 | 181.73—184.92 | 183.06 | 181.46—184.65 | 172.36 | 170.61—174.11 |
| 430 | 193.03 | 192.29—193.77 | 193.34 | 192.52—194.16 | 193.22 | 192.46—193.97 | 181.93 | 181.01—182.85 |
| 470 | 195.59 | 194.85—196.33 | 196.14 | 195.28—197.00 | 196.03 | 195.15—196.91 | 184.76 | 183.80—185.72 |

TABLE 17
BNC FOR K-GA AND OPTIMAL APPROACH IN CD SCENARIO

| Small Cell Density | Backhaul Network Capacity (Gbps) | | | | Gap between Optimal and K-GA (%) |
|---|---|---|---|---|---|
| | Optimal | | K-GA | | |
| | Mean | 95% CI | Mean | 95% CI | |
| 310 | 148.67 | 146.59—150.75 | 146.99 | 144.88—149.10 | 1.13% |
| 350 | 167.70 | 165.57—169.83 | 165.75 | 163.60—167.91 | 1.16% |
| 390 | 189.08 | 187.57—190.59 | 186.81 | 185.21—188.42 | 1.20% |
| 430 | 198.76 | 198.05—199.47 | 196.54 | 195.74—197.34 | 1.12% |
| 470 | 201.46 | 200.72—202.20 | 199.11 | 198.31—199.91 | 1.17% |

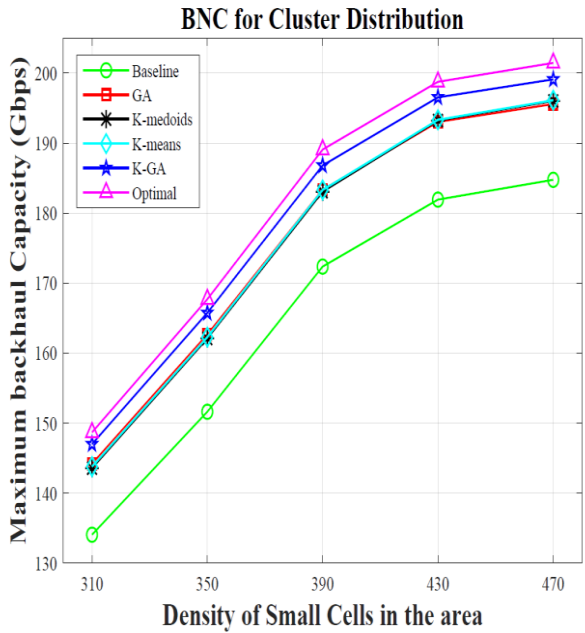

**FIGURE 17.** Backhaul Network Capacity in CD scenario

K-GA provides results within *2%* of the optimal value for all node densities in the cluster-based distribution scenario.

## IX. CONCLUSION AND FUTURE WORK

5G ultra-dense networks are necessary to achieve the high data rates promised by 5G. Backhaul optimization is a major challenge for accomplishing the performance objectives for such networks. Our backhaul network model uses a multi-hop distributed architecture similar to 3GPP's IAB architecture in Release 16. We examined the use of IAB as multi-hop networks delivering access traffic to the core.

We addressed the gateway location problem of selecting the best locations for gateways to ensure an effective backhaul network design. Our work shows that efficiently locating gateways improves the backhaul network capacity. Our new K-GA algorithm combines the simplicity of *K*-means with the evolutionary improvement in the genetic algorithm.

We also formulated the GLP as an integer linear program to obtain optimal gateway locations to allow us to analyze the quality of the K-GA solutions by comparison with the exact solution. The K-GA heuristic solves the GLP very quickly with a small percentage deviation from the optimal solution. We also compared the performance of our proposed heuristic with well-known heuristic techniques such as *K*-means algorithm, *K*-medoids algorithm, genetic algorithm, combination of *K*-medoids with genetic algorithm, and a baseline approach.

We tested K-GA in three different small cell distribution scenarios: uniform, bivariate Gaussian, and cluster distributions. K-GA provides better solutions in all three distribution environments compared to other approaches. K-GA saves on average *95%* of execution time compared to the optimal approach and provides ANH and BNC within *2%* of optimal.

There are several possible future research directions for extending and enhancing the work presented in this paper:

- Many fully loaded small cells and few gateways may cause congestion at gateways by aggregating small cell access traffic. This can be avoided by introducing a dynamic gateway allocation scheme. In case of congestion, such a scheme could add one or more gateways to the network and re-assign small cells among all gateways based on the minimum average number of hops. This would alleviate congestion in the backhaul network of 5G ultra dense networks.
- Mm-wave is a promising solution for backhaul connections to support the high data rate traffic demands in ultra-dense networks. However, mm-wave communications have short range due to high path loss. A comprehensive study is needed on dynamic beamforming
 as it can be useful in finding the optimum GW locations by decreasing the average number of hops.
- A scenario can be explored where small cells adjust their range to match different user distributions while optimizing the backhaul capacity and latency. Such a study would require a comprehensive probabilistic access traffic distribution.
- The possibility of deploying mobile GWs should be investigated.
- It will be interesting to explore a scenario where the traffic of the small cells exceeds the capacity of the gateways.

**REFERENCES**

[1] J. Thompson *et al*., "5G Wireless Communication Systems: Prospects and Challenges" *IEEE Communications Magazine*, vol. 52(2), pp. 62–64, 2014.

[2] N.H.M. Adnan, I. M. Rafiqul, and A. H. M. Z. Alam, "Massive MIMO for Fifth Generation (5G): Opportunities and Challenges," in proc. *International Conference on Computer and Communication Engineering (ICCCE),* Kuala Lumpur, pp. 47-52, 2016.

[3] T. S. Rappaport *et al*., "Millimeter Wave Mobile Communications for 5G Cellular: It Will Work!," *IEEE Access*, vol. 1, pp. 335-349, 2013.

[4] N. Bhushan *et al*., "Network Densification: The Dominant Theme for Wireless Evolution into 5G," *IEEE Communications Magazine*, vol. 52(2), pp. 82–89, 2014.

[5] M. Kamel, W. Hamouda, and A. Youssef, "Ultra-Dense Networks: A Survey," *IEEE Communications Surveys & Tutorials*, vol. 18(4), pp. 2522–2545, 2016.

[6] W. Feng, Y. Li, D. Jin, L. Su, "Millimeter-Wave Backhaul for 5G Networks: Challenges and Solutions," *Sensors*, vol. 16(6), 2016.

[7] 3GPP Release 16, [Online]. Available: https://www.3gpp.org/release-16.

[8] 3GPP TR 38.874, *NR; Study on Integrated Access and Backhaul*, [Online]. Available: https://www.3gpp.org/release-16.

[9] X. Ge, S. Tu, G. Mao, C. Wang, and T. Han, "5G Ultra-Dense Cellular Networks," *IEEE Wireless Communications*, vol. 23(1), pp. 72–79, 2016.

[10] P. Cappanera, and M. Nonato, "The Gateway Location Problem: Assessing the Impact of Candidate Site Selection Policies," *Discrete Applied Mathematics*, vol. 165, pp. 96–111, 2014.

[11] Ansari, Abdollah, and Azuraliza Abu Bakar. "A Comparative Study of Three Artificial Intelligence Techniques: Genetic Algorithm, Neural Network, and Fuzzy Logic, on Scheduling Problem," in proc. 4[th] *IEEE International Conference on Artificial Intelligence with Applications in Engineering and Technology*, pp. 31–36, 2014

[12] C.M. Bishop, *Pattern Recognition and Machine Learning*, Springer, New York, NY, 2006.

[13] C. Sammut, and G.I. Webb, *Encyclopedia of Machine Learning*, Springer, New York, NY, 2011.

[14] E.W. Dijkstra, "A Note on Two Problems in Connection with Graphs," *Numerische Mathematik*, vol. 1, pp. 269–271, 1959.

[15] M. Raithatha, A.U. Chaudhry, R.H.M. Hafez, and J.W. Chinneck, "Locating Gateways for Maximizing Backhaul Network Capacity of 5G Ultra-Dense Networks," in proc. *Wireless Telecommunications Symposium (WTS)*, Washington, DC, 2020.

[16] J. Reese, "Solution methods for the p-median problem: an annotated bibliography," *Networks*, vol. 48(3), pp. 125–142, 2006.

[17] S. Hakimi, "Optimum Locations of Switching Centers and the Absolute Centers and Medians of a Graph," *Operations Research*, vol. 12(3), pp. 450–459, 1964.

[18] D. W. Matula, and R. Kolde, "Efficient Multi-Median Location in Acyclic Networks," presented at *ORSA-TIMS meeting*, Miami, FL, 1976.

[19] O. Kariv and S.L. Hakimi, "An algorithmic approach to network location problems, Part II: p-medians", *SIAM Journal on Applied Mathematics*, vol. 37, pp. 539-560, 1979.

[20] W.L. Hsu, "The Distance-Domination Numbers of Trees," *Operations Research Letters*, pp. 96-100, 1982.

[21] A. Tamir, "An O(pn$^2$) Algorithm for the p-median and Related Problems on Tree Graphs," *Operations Research Letters*, Vol 19(2), pp. 59-64, 1996.

[22] R. Benkoczi and B. Bhattacharya, "A New Template for Solving p-median Problems for Trees in Sub-Quadratic Time," *Lecture Notes in Computer Science*, vol. 3669, pp. 271–282, Springer, Berlin, 2005.

[23] S. D. de S. Silva, M. G. F. Costa, and C. F. F. Costa Filho, "Customized Genetic Algorithm for Facility Allocation using p-median," in proc. *Federated Conference on Computer Science and Information Systems (FedCSIS)*, Leipzig, Germany, pp. 165-169, 2019.

[24] L. Haoze, "Research on Artificial Intelligence Optimization Based on Genetic Algorithm", *Light Industry Science and Technology*, pp.77-79, 2012.

[25] E. Wirsansky, "Hands-on Genetic Algorithms with Python : Applying Genetic Algorithms to Solve Real-World Deep Learning and Artificial Intelligence Problems" in *Hands-on Genetic Algorithms with Python*, Packt Publishing Limited, 2020.

[26] C. Guoliang, "Genetic Algorithm and Its Application", *Posts and Telecommunications Press*, 1999.

[27] J. Holland, "Concerning Efficient Adaptive Systems," in *Yovits,M.C.,Eds.,Self -Organizing Systems*, pp. 215 -230, 1962.

[28] F. Buontempo, "Genetic Algorithms and Machine Learning for Programmers : Create AI Models and Evolve Solutions," *The Pragmatic Bookshelf*, 2019.

[29] A. Kapoor, "Hands-On Artificial Intelligence for IoT: Expert machine learning and Deep Learning Techniques for Developing Smarter IoT Systems," *Packet Publishing Limited*, Birmingham, 2019.

[30] B. K. Ambati, J. Ambati, and M. M. Mokhtar, "Heuristic Combinatorial Optimization by Simulated Darwinian Evolution: A Polynomial Time Algorithm for the Traveling Salesman Problem," *Biological Cybernetics*, vol. 65(1), pp. 31–35, 1991.

[31] F. G. Lobo, D. E. Goldberg, and M. Pelikan, "Time complexity of Genetic Algorithms on Exponentially Scaled Problems," in proc. *Genetic and Evolutionary Computation Conference (GECCO '00)*, pp. 151–158, Las Vegas, Nevada, USA, 2000.

[32] K. Deb *et al*. "A Fast and Elitist Multi objective Genetic Algorithm: NSGA-II." *IEEE Transactions on Evolutionary Computation*, vol. 6(2), pp. 182–97, 2002.

[33] K. Krishna and M. Narasimha Murty. "Genetic K-Means Algorithm." *IEEE Transactions on Systems, Man, and Cybernetics, Part B (Cybernetics)*, vol. 29(3), pp. 433–39, 1999.

[34] C.W. Tsai *et al*. "A High-Performance Genetic Algorithm: Using Traveling Salesman Problem as a Case," *The Scientific World Journal*, vol. 2014, pp. 178621–178621, 2014.

[35] S. Yaram, "Machine Learning Algorithms for Document Clustering and Fraud Detection," in proc. *International Conference on Data Science and Engineering (ICDSE)*, Cochin, pp. 1-6, 2016.

[36] A.P. Tindi, R. Gernowo, and O.D. Nurhayati, "Machine learning: Fisher fund classification using neural network and particle swarm optimization," in proc. *Information and Communications Technology (ICOIACT)*, pp. 315-320, 2018.

[37] J. Xie, and S. Jiang, "A Simple and Fast Algorithm for Global K-means Clustering," in proc. *Second International Workshop on Education Technology and Computer Science*, Wuhan, pp. 36-40, 2010.

[38] K.A. Abdul Nazeer, and M.P. Sebastian, "Improving the Accuracy and Efficiency of the k-means Clustering Algorithm," in proc. *World Congress on Engineering (WCE)*, London, UK, 2009.

[39] L. Kaufman, and P.J. Rousseeuw, *Finding Groups in Data: An Introduction to Cluster Analysis*, Hoboken, NJ: John Wiley & Sons, Inc., 2009.

[40] R.T. Ng, and J. Han, "Efficient and Effective Clustering Methods for Spatial Data Mining," in proc. 20[th] *International Conference on Very Large Databases*, Santiago, Chile, 1994.

[41] N. Makariye, "Towards Shortest Path Computation Using Dijkstra Algorithm," in proc. *International Conference on IoT and Application (ICIOT)*, Nagapattinam, 2017.

[42] J. Donald, "A Note on Dijkstra's Shortest Path Algorithm," *Journal of the ACM*, vol. 20(3), pp. 385–88, 1973.

[43] H. Ortega-Arranz *et al*. "The Shortest-Path Problem : Analysis and Comparison of Methods," *Morgan & Claypool*, 2014.

[44] A. Madkour *et al*. "A Survey of Shortest-Path Algorithms," *arXiv.org, Cornell University Library*, 2017.

[45] M. Barbehenn, "A Note on the Complexity of Dijkstra's Algorithm for Graphs with Weighted Vertices," *IEEE Transactions on Computers*, vol. 47(2), pp. 263-, 1998.

[46] X. Ge, H. Cheng, M. Guizani, and T. Han, "5G Wireless Backhaul Networks: Challenges and Research Advances," *IEEE Network*, vol. 28(6), pp. 6-11, 2014.

[47] J. Lun, D. Grace, A. Burr, Y. Han, K. Leppanen, and T. Cai, "Millimeter Wave Backhaul/Fronthaul Deployments for Ultra-Dense Outdoor Small Cells," in proc. *IEEE Wireless Communications and Networking Conference workshops(WCNCW)*, pp. 187-192, Doha, 2016.

[48] Z. Gao, L. Dai, D. Mi, Z. Wang, M. A. Imran, and M. Z. Shakir, "Mm-wave Massive-MIMO-Based Wireless Backhaul for the 5G Ultra-Dense Network," *IEEE Wireless Communications*, vol. 22(5), pp. 13-21, 2015.

[49] C. Dehos, J. L. González, A. D. Domenico, D. Kténas, and L. Dussopt, "Millimeter-Wave Access and Backhauling: The Solution to the Exponential Data Traffic Increase in 5G Mobile Communications Systems," *IEEE Communications Magazine*, vol. 52(9), pp. 88-95, 2014.

[50] J.Y. Lai, W. Wu, and Y.T. Su, "Resource Allocation and Node Placement in Multi-Hop Heterogeneous Integrated-Access-and-Backhaul Networks," *IEEE Access*, vol. 8, pp. 122937–122958, 2020.

[51] A.L. Rezaabad, H. Beyranvand, J.A. Salehi, and M. Maier, "Ultra-Dense 5G Small Cell Deployment for Fiber and Wireless Backhaul-Aware Infrastructures," *IEEE Transactions on Vehicular Technology*, vol. 67, no. 12, pp. 12231–12243, Dec. 2018.

[52] Nidhi and A. Mihovska, "Small Cell Deployment Challenges in Ultra dense Networks: Architecture and Resource Management," in proc. *12th International Symposium on Communication Systems, Networks and Digital Signal Processing (CSNDSP)*, Porto, Portugal, pp. 1–6, 2020.

[53] J.F. Valenzuela-Valdés, Á. Palomares, J.C. González-Macías, A. Valenzuela-Valdés, P. Padilla, and F. Luna-Valero, "On the Ultra-Dense Small Cell Deployment for 5G Networks," in proc. *IEEE 5G World Forum (5GWF)*, Silicon Valley, CA, pp. 369–372, 2018.

[54] Y. Zhang, J. Liu, M. Sheng, Z. Xie, and J. Li, "Wireless Backhaul: Intrinsic Bottleneck of Ultra-Dense Networks?" in proc. *IEEE Global Communications Conference (GLOBECOM)*, pp. 1–6 Waikoloa, HI, USA, 2019.

[55] M.S. Haroon *et al.*, "Interference Management in Ultra-Dense 5G Networks With Excessive Drone Usage," *IEEE Access*, vol. 8, pp. 102155–102164, 2020.

[56] J. Liu, M. Sheng, L. Liu, and J. Li, "Interference Management in Ultra-Dense Networks: Challenges and Approaches," *IEEE Network*, vol. 31(6), pp. 70-77, 2017.

[57] D. Zhang and X. Tian, "Overview on Interference Management Technology for Ultra-Dense Network," *Open Access Library Journal*, vol. 5, pp. 1–14, 2018.

[58] A.U. Chaudhry, N. Jacob, D. George, and R.H.M. Hafez, "On the Interference Range of Small Cells in the Wireless Backhaul of 5G Ultra-Dense Networks," in proc. *Wireless Telecommunications Symposium (WTS)*, Washington, DC, USA, 2020.

[59] A.U. Chaudhry, N. Jacob, D. George, and R.H.M. Hafez, "On Evaluating Independent Set Heuristics for Wireless Backhaul Network Capacity of 5G Ultra-Dense Networks," in proc. *International Symposium on Networks, Computers and Communications (ISNCC)*, Montreal, QC, 2020.

[60] Z. Liu, X. Chen, Y. Chen, and Z. Li, "Deep Reinforcement Learning Based Dynamic Resource Allocation in 5G Ultra-Dense Networks," in proc. *IEEE International Conference on Smart Internet of Things (SmartIoT)*, Tianjin, China, pp. 168–174, 2019.

[61] W. Li *et al.*, "Energy Efficiency Maximization Oriented Resource Allocation in 5G Ultra-Dense Network: Centralized and Distributed Algorithms," *Computer Communications*, vol. 130, pp. 10–19, 2018.

[62] M. Polese *et al.*, "Integrated Access and Backhaul in 5G mmWave Networks: Potential and Challenges," *IEEE Communications Magazine*, vol. 58(3), pp. 62–68, March 2020.

[63] N. Palizban, "Millimeter Wave Small Cell Network Planning for Outdoor Line-of-Sight Coverage," *M.A.Sc. thesis, Systems and Computer Engineering, Carleton University*, Ottawa, Canada, 2017.

[64] P. Huang and K. Psounis, "Optimal Backhauling for Dense Small-Cell Deployments using mmWave Links," *Computer Communications*, vol. 138, pp. 32-44, 2019.

[65] X. Ge, L. Pan, S. Tu, H. Chen, and C. Wang, "Wireless Backhaul Capacity of 5G Ultra-Dense Cellular Networks," in proc. *IEEE 84th Vehicular Technology Conference (VTC-Fall)*, Montreal, 2016.

[66] X. Ge, S. Tu, G. Mao, V. Lau, and L. Pan, "Cost Efficiency Optimization of 5G Wireless Backhaul Networks," *IEEE Transactions on Mobile Computing*, vol. 18(12), pp. 2796-2810, 2019.

[67] H. Wang, S. Chen, M. Ai, and H. Xu, "Localized Mobility Management for 5G Ultra Dense Network," *IEEE Transactions on Vehicular Technology*, vol. 66, no. 9, pp. 8535-8552, Sept. 2017.

[68] S.K. Ghosh and S.C. Ghosh, "Analyzing Handover Performances of Mobility Management Protocols in Ultra-dense Networks," *Journal of Network and Systems Management*, vol. 28, pp. 1427–1452, 2020.

[69] J. Zhang, J. Feng, C. Liu, X. Hong, X. Zhang, and W. Wang, "Mobility Enhancement and Performance Evaluation for 5G Ultra Dense Networks," in proc. *IEEE Wireless Communications and Networking Conference (WCNC)*, New Orleans, LA, pp. 1793–1798, 2015.

[70] Y. Cao, L. Zhao, Y. Shi, and J. Liu, "Gateway Placement for Reliability Optimization in 5G-Satellite Hybrid Networks" in proc. *International Conference on Computing, Networking and Communications (ICNC)*, Maui, 2018.

[71] J. Nie, D. Li, Y. Han, W. Fu, and G. Zhang, "The Method of Multiple Surface Gateways Positioning in UWSNs," in proc. *6th International Conference on Wireless Communications Networking and Mobile Computing (WiCOM)*, Chengdu, China, 2010.

[72] M. Awadallah, and H. Aisha, "A Genetic Approach for Gateway Placement in Wireless Mesh Networks," *International Journal of Computer Science and Network Security*, vol. 15(7), pp. 11-19, 2015.

[73] T. Mahmoud, M. Girgis, B. Abdullatif, and A. Rabie, "Solving the Wireless Mesh Network Design Problem using Genetic Algorithm and Simulated Annealing Optimization Methods," *International Journal of Computer Applications*, vol. 96(11), 2014.

[74] A.M. Ahmed, A.H.A. Hashim, and W.H. Hassan, "Investigation of Gateway Placement Optimization Approaches in Wireless Mesh Networks Using Genetic Algorithms," in proc. *International Conference on Computer and Communication Engineering*, Kuala Lumpur, 2014.

[75] W.L. Martinez and A.R. Martinez, *Computational statistics handbook with MATLAB*, Chapman & Hall/CRC, Boca Raton, FL, USA, 2002.

[76] IBM ILOG CPLEX Studio IDE, Version: 12.10.0.201911271611, Build id: 201911271611, Licensed Materials - Property of IBM Corp. © Copyright IBM Corporation and other(s) 1998, 2019.

[77] N. Mladenović, J. Brimberg, P. Hansen, and A. Moreno-Pérez, "The p-median Problem: A Survey of Metaheuristic Approaches," *European Journal of Operational Research*, vol. 179(3), pp. 927-939, 2007.

[78] S. Silva, M. Costa, and C. Filho, "Customized Genetic Algorithm for Facility Allocation using p-median," *Federated Conference on Computer Science and Information Systems (FedCSIS)*, Leipzig, Germany, pp. 165-169, 2019.

[79] J. Fathali, "A Genetic Algorithm for the p-median Problem with pos/neg Weights," *Applied Mathematics and Computation*, vol. 183, pp. 1071-1083, 2006.

[80] O. Alp, E. Erkut, and Z. Drezner, "An Efficient Genetic Algorithm for the p-Median Problem," *Annals of Operations Research* vol. 122, pp. 21–42, 2003.

[81] E.S. Correa, M. Steiner, A. Freitas *et al.*, "A Genetic Algorithm for Solving a Capacitated p-Median Problem," *Numerical Algorithms*, vol. 35, pp. 373–388, 2004.

[82] P. Rebreyend, L. Lemarchand, and R. Euler, "A Computational Comparison of Different Algorithms for Very Large p-median Problems," in *Evolutionary Computation in Combinatorial Optimization (EvoCOP), Lecture Notes in Computer Science*, vol. 9026, Springer, Copenhagen, 2015.

[83] Z. Drezner, J. Brimberg, N. Mladenović, and S. Salhi, "New Heuristic Algorithms for Solving the Planar p-median Problem," *Computers & Operations Research*, vol. 62, pp. 296-304, 2015.

[84] A. Chehouri, R. Younes, J. Khoder *et al.*, "A Selection Process for Genetic Algorithm Using Clustering Analysis," *Algorithms* vol. 10, no 4:123, 2017.

[85] S. Kapil, M. Chawla, and M.D. Ansari, "On K-means Data Clustering Algorithm with Genetic Algorithm," in proc. *Fourth International Conference on Parallel, Distributed and Grid Computing (PDGC)*, pp. 202–206, 2016.

[86] M. Dunham *et al.*, "Clustering", in *Data Mining: Introductory and Advanced Topics*, Upper Saddle River, NJ: Prentice Hall, 2002.

[87] S.N. Sivanandam, and S.N. Deepa, "Terminologies and Operators of GA" in *Introduction to Genetic Algorithms*, Springer Berlin Heidelberg, 2008.

[88] H.P. Keeler, "Notes on the Poisson Point Process," *Weierstrass Institute*, Berlin, Germany, Tech. Rep., 2016.

[89] J.G. Andrews, T. Bai, M.N. Kulkarni *et al.*, "Modeling and Analyzing Millimeter Wave Cellular Systems," *IEEE Transactions on Communications*, vol. 65(1), pp. 403–430, 2017.

[90] M.R. Akdeniz, Y. Liu, M.K. Samimi *et al.*, "Millimeter Wave Channel Modeling and Cellular Capacity Evaluation," *IEEE Journal on Selected Areas in Communications*, vol. 32(6), pp. 1164–1179, 2014.

[91] A.U. Chaudhry, R.H.M. Hafez, O. Aboul-Magd, and S.A. Mahmoud, "Throughput Improvement in Multi-Radio Multi-Channel 802.11a-Based Wireless Mesh Networks," in proc. *IEEE Global Telecommunications Conference* (*GLOBECOM*), Miami, FL, 2010.

[92] V.H. Pascal, and I. Lustig, "The OPL Optimization Programming Language," in *Cambridge, Mass: MIT Press*, 1999.

[93] The MathWorks Inc., "MATLAB® Reference Guide," Natick, MA, USA, 1992.

[94] M. Jaber, M.A. Imran, R. Tafazolli, and A. Tukmanov, "5G Backhaul Challenges and Emerging Research Directions: A Survey," *IEEE Access*, vol.4, pp. 1743–1766, 2016.

[95] 5G Americas and Small Cell Forum, "Deployment Plans and Business Drivers for a Dense HetNet: SCF Operator Survey," Tech. Rep. 194.10.01, Dec. 2017, [Online]. Available: https://scf.io/en/download.php?doc=194.

[96] 5G Americas and Small Cell Forum, "5G Small Cell Architecture and Product Definitions Configuration and Specifications for Companies Deploying Small Cells 2020–2025," Tech. Rep. 238.10.01, Jul. 2020, [Online]. Available: https://scf.io/en/download.php?doc=238.

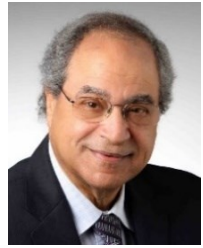

**ROSHDY H.M. HAFEZ** (M'86–LSM'19) obtained the Ph.D. degree from Carleton University in Electrical and Computer Engineering. He has a long research career in "wireless systems and networks". He supervised more than 60 Master's and Ph.D.'s research theses in the same area. He acted as a technical consultant to many companies and government agencies in Canada, USA and Europe. He taught many professional courses on detailed issues related to international wireless standards from GSM to 5G.

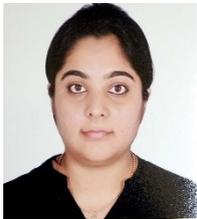

**MITAL RAITHATHA** received Bachelor of Engineering in Electronics and Communication engineering from Gujarat Technological University, India in 2014. She received Master of Applied Science in Electrical and Computer engineering from Carleton University in 2020. She worked in many Information Technology companies for 3 years in India. Her research interests include 5G networks, machine learning, artificial intelligence, and wireless network planning and optimization.

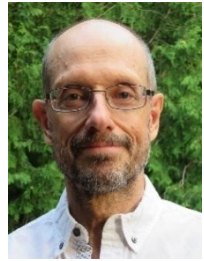

**JOHN W. CHINNECK** received the Ph.D. degree in Systems Design Engineering from the University of Waterloo in 1983. He researches the development and application of optimization algorithms. He served as the Editor-in-Chief of *The INFORMS Journal on Computing* from 2007 to 2012, and as the Chair of the INFORMS Computing Society from 2006 to 2007. He has received Research Achievement Awards from Carleton University, and the Award of Merit from the Canadian Operational Research Society, among other awards.

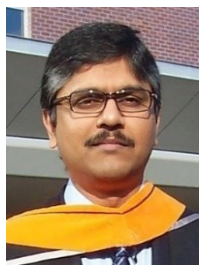

**AIZAZ U. CHAUDHRY** (M'10–SM'20) received the B.Sc. degree in Electrical Engineering from the University of Engineering and Technology Lahore in 1999. He received the M.A.Sc. and Ph.D. degrees in Electrical and Computer Engineering from Carleton University in 2010 and 2015, respectively. Aizaz is a Research Associate with the Department of Systems and Computer Engineering, Carleton University. Previously, he worked as an NSERC Postdoctoral Research Fellow at Communications Research Centre Canada. His research interests include the application of machine learning and optimization in wireless networks.